\documentclass[aps,twocolumn,longbibliography,superscriptaddress,nofootinbib]{revtex4-1}
\usepackage{amsfonts}
\usepackage{libertine}
\usepackage{amsmath,amssymb,epsfig,color}
\usepackage{float}
\usepackage{amsmath}
\usepackage{amssymb}
\usepackage{url}
\usepackage{hyperref}
\usepackage{graphicx}
\usepackage{dcolumn}
\usepackage{bbold}
\usepackage{comment}
\DeclareMathOperator{\sgn}{sgn}
\begin{document}

\title{Anomalous Josephson diode effect in superconducting multilayers}

\author{A. S. Osin}
\affiliation{Racah Institute of Physics, Hebrew University of Jerusalem, Jerusalem 91904, Israel}

\author{Alex Levchenko}
\affiliation{Department of Physics, University of Wisconsin--Madison, Madison, Wisconsin 53706, USA}

\author{Maxim Khodas}
\affiliation{Racah Institute of Physics, Hebrew University of Jerusalem, Jerusalem 91904, Israel}

\date{June 17, 2024}

\begin{abstract}
In this study, we explore the Josephson current-phase relation within a planar diffuse tunneling superconducting multilayer junction subjected to a parallel magnetic field. Our investigation involves computing the supercurrent associated with a fixed jump in the phase of the order parameter at each of the two insulating interfaces, allowing us to derive the current-phase relation for the junction. Employing perturbation theory in junction conductance, we determine both the first and second harmonics of the current-phase relation under specific magnetic field conditions. Notably, the presence of a strong spin-orbit interaction in the middle region of the junction introduces an anomalous Josephson effect. The interplay between spin-orbit and Zeeman interactions results in the emergence of an effective vector potential. This specific characteristic induces a phase shift in each harmonic of the current-phase relation without altering the overall shape of the relation.
The mechanism for the Josephson diode effect is discussed for disordered junctions of multiband superconductors.   
\end{abstract}
\maketitle

\section{Introduction}

The Josephson junction is the key element of superconducting circuits where the superconducting phase, $\varphi$ retains coherence on a macroscopic scale. The Josephson junction forms a weak link between the two superconductors \cite{Golubov2004}. The weak link can be a constriction, insulating (I) tunnel barrier, normal (N) metal, ferromagnet (F), semiconductor (Sm), or other superconductor (S). Regardless of the particular realization, the defining property of the Josephson junction is the current-phase relation, $J=J(\Delta\varphi)$ between the change of the phase of the order parameter (OP) $\Delta \varphi$ across the junction and the Josephson current $J$.

As the phase is defined up to a multiple of $2 \pi$ current-phase relation is periodic, $J(\Delta\varphi \pm 2 \pi )=J(\Delta\varphi)$.
In addition, both the time-reversal symmetry ($\mathcal{T}$) and other parity and/or parity-like unitary symmetries ($\mathcal{I}$) inverting the current direction imply $J(\Delta\varphi) = -J(-\Delta\varphi)$. 
As a result, when at least one of these symmetries is present, the current-phase relation satisfies $J(\Delta \varphi=0) = 0$.
The simplest current-phase relation satisfying these requirements, e.g., in the SIS junction, is $J= J_c \sin (\Delta \varphi)$, where $J_c$ is the critical current. In contrast, if all of the above symmetries are broken, $J(\Delta \varphi=0) \neq 0$. In this, less common scenario the Josephson junction is said to exhibit the anomalous Josephson effect (AJE).
Previously, the AJE has been explored in junctions formed by the unconventional superconductors \cite{Geshkenbein1986,Yip1995,Sigrist1998,Kashiwaya2000}.

The simplest model of the current-phase relation of Josephson junction in AJE regime is $J= J_c \sin (\Delta \varphi + \varphi_0)$.
Hence, the Josephson junctions exhibiting AJE are characterized by a $\varphi_0 \neq 0$, and are referred to as $\varphi_0$-Josephson junction \cite{Buzdin2008}.
In superconductor-ferromagnet (SF) structures with broken $\mathcal{T}$ symmetry for some thicknesses of the ferromagnet, the exchange energy might turn $\varphi_0$ to $\pi$ (see Refs. \cite{Buzdin1982,Ryazanov2001,Buzdin2005,Braude2007,Houzet2007,Gingrich2016}). 
This means that in the ground state of such a $\pi$ junction the OP changes the sign across the junction.

To highlight the significance of the AJE consider the isolated superconducting loop with a  $\varphi_0$-Josephson junction threaded by the Aharonov-Bohm flux, $\Phi=\Phi_{\text{ext}}+\mathcal{L}J$, where $\Phi_{\text{ext}}$ is the external magnetic flux, and $\mathcal{L}$ is the inductance of the loop.
The fluxoid quantization gives $\Delta \varphi = 2 \pi \Phi/\Phi_0$ up to a multiple of $2 \pi$ with the flux quantum $\Phi_0 =  \pi\hbar/e c$. 
Hence, the current in the loop in the simplest case, 
$J_c \sin (2 \pi \Phi/\Phi_0 + \varphi_0)$.
At $\Phi_{\text{ext}}=0$ both the current and the flux vanish in the normal Josephson junction with $\varphi_0=0$. 
In contrast, finite $\varphi_0$ in AJE implies finite current in the loop.
From now on we use units $\hbar=k_B=c=1$.

Recently, the AJE in the planar superconductor-normal-superconductor (SNS) structures has been observed experimentally \cite{Szombati2016,Assouline2019,Mayer2020,Strambini2020,Guarcello2020,Idzuchi2021,Baumgartner2022,Margineda2023} and  studied theoretically \cite{Bergeret2015,Konschelle2015,Rasmussen2016,Silaev2017,Fyhn2020,Hasan2022}.
The Rashba-type spin-orbit interaction breaking the in-plane mirror symmetry along with the magnetic field breaking the $\mathcal{T}$ symmetry lead a finite phase shift, $\varphi_0$ as well as an asymmetry in the Fraunhofer diffraction pattern when a finite magnetic flux threads the junction of a finite width.
In all such cases the AJE is sensitive to the length of the $N$ region $L_N$ as well as to the presence of the tunnel barriers between the normal and superconducting regions.
For the clean and transparent SNS structures $\varphi_0 \propto L^3_N$ \cite{Hasan2022} while in the case of SINIS structures with the insulating low-transparency $I$ regions $\varphi_0 \propto L_N$ \cite{Konschelle2015}.
For the disordered $N$ region,  $\varphi_0 \propto L^{2}_{N}$ for long junctions, and  $\varphi_0 \propto L_N$ for short $N$ regions with tunnel barriers, and for transparent and short junction $\varphi_0 \propto L_N^3$ as in the clean case \cite{Bergeret2015}.

In this paper, we study the current-phase relation and AJE of the disordered multilayer SIS$^\prime$IS structures in the planar geometry. 
The superconducting S regions and S$^\prime$ region have generically different critical temperatures denoted as $T_c$ and $T_c'$, respectively. Correspondingly, the OPs, $\Delta(T) \neq \Delta'(T)$ are finite for considered temperatures, $T < \min\{T_c,T_c'\}$. All the regions are assumed to be in the diffusive limit with the typical mean-free path being the shortest length scale in the problem. The insulating tunnel barriers are assumed to have a low transparency. We find that under the above assumptions spin-orbit interaction in combination with the in-plane magnetic field gives rise to the AJE with $\varphi_0$ scaling linearly with the system size, $L$.

As shown in Ref.~\cite{Houzet2015} the in-plane magnetic field, $\mathbf{B}$ generates the emergent vector potential, $\mathbf{A}_{\mathrm{eff}}$ in the spin-orbit-coupled superconductor. From this result we demonstrate that $\varphi_0 = 2 e A_{\mathrm{eff}} L$, is the anomalous phase at least for sufficiently low transparencies of the I regions.
This conclusion holds even if the harmonics higher than the first one are retained in the current-phase relation.
The linear scaling of $\varphi_0$ has also been obtained for junctions with the weak link formed due to the geometric constriction \cite{Hasan2022}.

Within the Ginzburg-Landau phenomenology, the emergent vector potential $\mathbf{A}_{\mathrm{eff}}$ can be understood as arising from the so-called Lifshitz invariants appearing in the free-energy functional \cite{Smidman2017}. 
These terms are quadratic in the OP and are linear in the OP gradients. In superconductors with the time-reversal symmetry, $\mathcal{T}$, Lifshitz invariants require breaking of $\mathcal{T}$ by an external field, typically an external magnetic field $\mathbf{B}$.
In addition, since gradients and $\mathbf{B}$ are polar and axial vectors, respectively, the inversion center has to be absent.
The existence and the algebraic form of the Lifshitz invariants depend on the point-group symmetry as tabulated in Ref.~\cite{Smidman2017}. 
Generally, the Lifshitz invariants inherit their structure from the spin-orbit-coupling Hamiltonian that is linear in momentum.

In the bulk, Lifshitz invariant can be gauged out by the transformation, $\varphi(\mathbf{r}) \rightarrow \varphi(\mathbf{r}) - 2 e \mathbf{A}_{\mathrm{eff}} \mathbf{r}$.
The resulting helical ground state carries no current.
Moreover, the critical current is isotropic.
In other words, no superconducting diode effect (SDE) arises.
It follows that the bulk SDE is finite in cases when gauging out of $\mathbf{A}_{\mathrm{eff}}$ is impossible. 
For instance, cubic and/or higher gradients terms in the free energy give rise to SDE \cite{Daido2022}.

The ballistic high transparency $\varphi_0$-Josephson junction exhibits SDE once the second or higher harmonics are present in the current-phase relation \cite{Baumgartner2022a,Baumgartner2022,Reinhardt2023}. 
The effective vector potential $\mathbf{A}_{\mathrm{eff}}$ is not the same for different ballistic channels.
As a result, it cannot be gauged out. 
Still, this does not imply a finite SDE.
The other important ingredient is the skewness of the current phase relation.
This is guaranteed if the current has second and/or higher harmonics. 

Motivated by these findings we pose the question whether one can have a universal SDE in the opposite limit of dirty $\varphi_0$-Josephson junction.
We demonstrate that the second harmonics is necessary for SDE exactly as in the ballistic case.
However, in addition more than one band is required.
In this paper we study in details both aspects of the SDE in dirty junctions.
To this end, we consider the minimal model of S$^\prime$ superconductor supporting a finite $A_{\mathrm{eff}}$.
We supplement the Usadel equations describing the OP in S and S$^\prime$ regions formulated for the Rashba spin-orbit-coupled superconductors in Ref.~\cite{Houzet2015} by the Kuprianov-Lukichev boundary conditions \cite{KL1987} properly extended to account for a finite $A_{\mathrm{eff}}$ \cite{Bergeret2015}.

We fix the current-carrying state by the
phase difference $\Delta\varphi$ of the OP just behind the interfaces, I in the S electrodes (see Fig.~\ref{fig:symmetries}). 
This parameter is directly measurable. 
However, it is more convenient initially to perform all the calculations in terms of the phase jump of the OP across each of the interfaces, $\delta\varphi$.
This jump is the same for both of the interfaces due to symmetry. 
The total phase jump of the OP $\Delta\varphi$ is the sum of the two jumps at the boundaries $\delta\varphi$ and the phase accumulation inside the S$^\prime$ electrode.

The paper is organized as follows.
In Sec.~\ref{sec:symmetry} we discuss the limitation imposed by symmetry on the model to yield AJE.
This allows us to formulate the minimal model of S$^\prime$ superconductor with an emerging effective vector potential. In Sec. \ref{sec:results} we present the main findings of this work. In particular, the solution for the current is given up to the second harmonic in the perturbation theory developed over the barrier transparency. The self-consistent results for the OP are also presented.  In Sec.~\ref{sec:UsadelBC} we lay out the set of Usadel equations, and describe the basic solution method. The most technical parts of the calculation are given in the Appendixes.  

\section{Symmetry considerations}\label{sec:symmetry}

\begin{figure}[t!]
    \includegraphics[width=0.95\columnwidth]{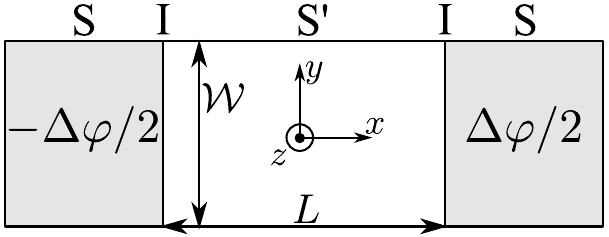}
    \caption{
    Quasi-2D  rectangular SIS$^\prime$IS junction lying in $xy$-plane.
    The insulating barriers I separate S regions (shaded) and S$^\prime$ region (white). The width and the length of the S$^\prime$ region is $\mathcal{W}$ and $L$, respectively. Symmetry operations of the SIS$^\prime$IS form a $D_{2h}$ point symmetry group with elements listed in Table~\ref{table:1}.
    Only those operations that exchange the two S regions, effectively change the sign of the phase difference, $\varphi \rightarrow - \varphi$ have to be broken for a finite AJE.
     }
    \label{fig:symmetries}
\end{figure}

We derive the necessary conditions for the AJE in a thin quasi-two-dimensional (quasi-2D) SIS$^\prime$IS Josephson junction (see Fig.~\ref{fig:symmetries}).
As such our arguments are similar to those of Ref.~\cite{Rasmussen2016}, where we apply the symmetry operations to the S$^\prime$  rather than the normal region.
Unlike in the planar geometry, supporting the Fraunhofer field dependence of the critical current, we have no perpendicular magnetic field in the S$^{\prime}$ region.
This extends the list of symmetry operations that effectively flip the phase $\Delta \varphi$ across the junction. 
For our purposes, it is sufficient to consider the operations acting on both orbital and spin degrees of freedom.
This is justified since the inversion operation, $\mathcal{I}$ has to be broken for AJE in any scenario. 
As a result, the spin and orbital motions are coupled by the atomic spin-orbit interaction.

As the supercurrent flow in our model is essentially one dimensional, we envision the S$^\prime$ region as a thin rectangular prism with a pair of opposite bases formed by the two insulating barriers denoted $A$ and $B$. 
The phases of the OP in the S region right next to the insulating barriers, $\pm \Delta \varphi/2$ are fixed by Aharonov-Bohm flux throughout this section.

We start with the S$^\prime$ which is fully isotropic and consider separately the symmetry breaking by the in-plane magnetic field, and the spin-orbit coupling. The OP S$^\prime$ is assumed to respect $\mathcal{T}$ and $\mathcal{I}$ symmetries. This is in cases with pseudo-scalar OP odd under $\mathcal{I}$ the Josephson current. 

For the isotropic S$^\prime$, the rectangular prism has a $D_{2h}$ point group. Some of the operations exchange the insulating barriers, $A \leftrightarrow B$ and some of them leave them unaffected.
Exchanging the barriers amounts to flipping the sign of the phase, $\Delta \varphi$. The $\mathcal{T}$ operation causes the same effect.
One way to see this is to note that the Aharonov-Bohm field is reversed by $\mathcal{T}$.

Table \ref{table:1} summarizes the properties of the elements of the point group, $D_{2h}$.
Among the elements of $D_{2h}$, rotations by $\pi$ around the $y$ axis, $C_{2y}$ as well as the rotation $C_{2z}$, the mirror $\sigma_{yz}$ in $yz$ plane effectively flips the phase difference, $\Delta\varphi$. 
The same is true for the inversion, $\mathcal{I}$ and $\mathcal{T}$.
Therefore, the unitary symmetries that have to broken for a finite AJE either extrinsically by the in-plane field or intrinsically by the spin-orbit couplings are $C_{2y,2z}$, $\sigma_{yz}$, $\mathcal{I}$.
The nonunitary symmetries $\mathcal{T}$, $\mathcal{T} C_{2x}$, $\mathcal{T} \sigma_{xy,xz}$ also flip $\Delta \varphi$ and have to be broken.

\begin{table}[t!]
\centering
\begin{tabular}{|c||c|c|c|c|c|c|c|c||c|} 
 \hline
   & $E$ & $C_{2x}$ & $C_{2y}$ & $C_{2z}$ & $\sigma_{xy}$ & $\sigma_{yz}$ & $\sigma_{xz}$ & $\mathcal{I}$ & $\mathcal{T}$\\  
 \hline
 $\Delta \varphi \rightarrow - \Delta \varphi$ & $+$ &
 $+$ & $-$ & $-$ & $+$ & $-$ & $+$ & $-$ & $-$ \\
 \hline
 \hline
 $B_x$ & $+$ & $+$ & $-$ & $-$ & $-$ &
 $+$ & $-$ &
 $+$ & $-$ \\
 \hline
 $B_y$ & $+$ & $-$ & $+$ & $-$ & $-$ &
 $-$ & $+$ &
 $+$ & $-$ \\
 \hline
 \hline
 $\alpha_R$ &  $\checkmark$ & $\text{\sffamily X}$ & $\text{\sffamily X}$ & $\checkmark$ & $\text{\sffamily X}$ & $\checkmark$ & $\checkmark$ & $\text{\sffamily X}$ & $\checkmark$ \\
 \hline
 $\beta_D$ & $\checkmark$ & $\checkmark$ & $\checkmark$ & $\checkmark$ & $\text{\sffamily X}$ & $\text{\sffamily X}$ & $\text{\sffamily X}$ & $\text{\sffamily X}$ & $\checkmark$ \\
 \hline
\end{tabular}
\caption{$D_{2h}$ symmetries label the columns. 
$C_{2t}$ is the rotation by $\pi$ around the $t$ axis,
$\sigma_{tt'}$ is the mirror in $tt'$ plane.
Second row marks by $+$ ($-$) the symmetries preserving (flipping) the sign of $\Delta \varphi$. The $x$ ($y$) components of $\mathbf{B}$  
are multiplied by $\pm 1$ in accordance with the entries in the third and fourth rows. The last two rows indicate which of the symmetries is preserved ($\checkmark$) or broken ($\text{\sffamily X}$) by Rashba ($\alpha_R$) and Dresselhaus ($\beta_D$) spin-orbit couplings.}
\label{table:1}
\end{table}

Before we analyze the effect of the two components of the field $\mathbf{B}$ let us state the general requirements on the system.
First, unless the S$^\prime$ breakes $\mathcal{T}$ spontaneously, the time reversal has to be broken extrinsically by the applied field(s). 
Second, as $\mathbf{B}$ is an axial vector, it is invariant under $\mathcal{I}$ and therefore the inversion symmetry has to broken intrinsically. Normally this causes the spin splitting of the electronic bands. Third, breaking of the $C_{2z}$ is not required for AJE because an in-plane field breaks it anyway. And, finally, the $\sigma_{xy}$ has to broken intrinsically, as otherwise $\sigma_{xy} \mathcal{T}$ is the symmetry even for finite in-plane $\mathbf{B}$. This means that Ising superconductors do not show AJE. 

We now consider the combined action of the in-plane magnetic field and the spin-orbit coupling. The two types of spin orbit we consider here are Rashba spin-orbit coupling, $H_{R}=\alpha_R (\sigma_{x} p_y -\sigma_{y} p_x)$, where  $p_{x,y}$ are the components of the in-plane momentum, and the Dresselhaus spin-orbit interaction of the form, $H_{D}=\beta_D (\sigma_{x} p_x -\sigma_{y} p_y)$.

As it follows from Table \ref{table:1} for $\mathbf{B} = \hat{x} B_x$ ($\mathbf{B} = \hat{y} B_y$) the Dresselhaus (Rashba) spin-orbit coupling causes the AJE, while Rashba (Dresselhaus) spin-orbit coupling does not.

\section{Summary of Results}\label{sec:results}

In this section we summarize and discuss our main findings.
The junction is modeled as an infinite quasi-two-dimensional (2D) strip of a transverse width, $\mathcal{W}$. The S$^\prime$ region has length, $L$, and occupies the region $|x|< L/2$ (see Fig.~\ref{fig:symmetries} for the illustration). The analysis is trivially generalized to the three-dimensional geometry with $\mathcal{W}$ replaced by the cross-section area.

The S (S$^\prime$) region is assumed to be an $s$-wave superconductor with an equilibrium gap $\Delta_0\ (\Delta'_0)$, the mean-free path $\ell\ (\ell')$, and the Fermi velocity $v_F\ (v'_F)$.
Here $T$ is the temperature.
The normal-state conductivity in the S region is $\sigma = 2 e^2 \nu_0 D$,
where $\nu_0$ is the density of states per spin, and $D= v_F \ell/2$ is the diffusion coefficient in a quasi-two-dimensional geometry.
Similar definitions apply for the S$^\prime$ region, where hereinafter the quantities referring to S and S$^\prime$ regions are not primed and primed, respectively.   

Each of the two regions is characterized by its respective coherence length $\xi$ and $\xi'$.
Here we adopt the operational definition $\xi = \sqrt{D/[2 \Delta_0(T)]}$, $\xi' = \sqrt{D'/[2 \Delta'_0(T)]}$ appearing naturally in our analysis. 
It should be noted that $\xi$ scales as $(T-T_c)^{-1/4}$, $T$ close to the critical temperature $T_c$. This scaling differs from the one expected from the Ginzburg-Landau theory, $\xi_{\text{GL}} \propto (T-T_c)^{-1/2}$. 
We later comment on this difference. 
Both regions are assumed to be in the diffusive limit, $\ell \ll \xi$, $\ell' \ll \xi'$, where the Usadel equations as formulated in Sec.~\ref{sec:UsadelBC} apply. 

We are considering the tunneling limit, which implies that the interface conductance, $G_{\mathrm{int}}$
of the interface is much lower than every other characteristic conductance
of the system~\footnote{In the paper \cite{Osin2021} the tunneling criterion
was also discussed. 
Near the critical temperature $T_{c}$ the coherence length $\xi$ should be replaced by $\xi_{\text{GL}}=\sqrt{\pi D/8\left(T_{c}-T\right)}\gg\xi$. 
This is considered in more details in Appendix~\ref{AppendixApplicability}.\label{Tunneling criterion footnote}}:
\begin{equation}
\alpha\! =\! \frac{g_{\text{int}}\xi}{\sigma }\ll 1, \quad\alpha^{\prime}\!=\!\frac{g_{\text{int}}\xi^{\prime}}{\sigma^{\prime}}\ll1,\quad
\alpha_{L}\!=\!\frac{g_{\text{int}}L}{\sigma^{\prime}}\ll 1,
\label{eq:AlphaIneq}
\end{equation}
where $g_{\mathrm{int}}= G_{\text{int}}/W$.
Here we present the results of the perturbation theory in small parameters of Eq.~\eqref{eq:AlphaIneq}.
In SIS junctions the perturbation theory in $\alpha \ll 1$ accounts for the second harmonics in the current-phase relation $\propto \alpha^2 \sin(2 \delta\varphi)$ \cite{Osin2021}. 

The absolute value and phase of the OP are even and odd function of the coordinate, $x$. Therefore, we specify these results in the S region for $x>L/2$. The results below are obtained by solving the Usadel equations along with the boundary conditions in S and S$^\prime$ regions in perturbation theory in the parameter $\alpha$ and $\alpha'$, respectively.
More specifically we look for the expressions for the absolute value of the OP, and its phase in the S region,
\begin{align}\label{eq:resS}
    |\Delta(x)| & = \Delta_0 + \alpha \Delta_{1}(x)
    \notag \\
    \varphi(x) & = eA_{\mathrm{eff}}L + \delta \varphi + \alpha \varphi_1(x)  
\end{align}
In the S$^\prime$ region we similarly have,
\begin{align}\label{eq:resS'}
    |\Delta'(x)| & = \Delta'_0 + \alpha' \Delta'_{1}(x)
    \notag \\
    \varphi'(x) & = 2 eA_{\mathrm{eff}} x  + \alpha' \varphi'_1(x)  \, .
\end{align}

In section \ref{sec:OPabs} we specify and illustrate the corrections, $\Delta_1$ and $\Delta'_1$ to the absolute value of the OP.
In the following Sec. \ref{sec:OPphase} we discuss and illustrate the corrections to the phase of the OP, $\varphi_1$ and $\varphi'_1$.
The corrections to the current are presented in Sec. \ref{sec:J}.
The formalism used to obtain these results is outlined in Sec.  \ref{sec:UsadelBC}.
The details of the derivations are relegated to Appendixes~\ref{AppendixPerturbation theory} and ~\ref{AppendixCPR}.
The expressions for the OP and the current for the long, $L \gg \xi'$ and short $L \ll \xi'$ S$^\prime$ regions are obtained in the Appendixes~\ref{AppendixLongJunction} and  ~\ref{AppendixShortJunction}.

\begin{figure}[t!]
    \includegraphics[width=0.45\textwidth]{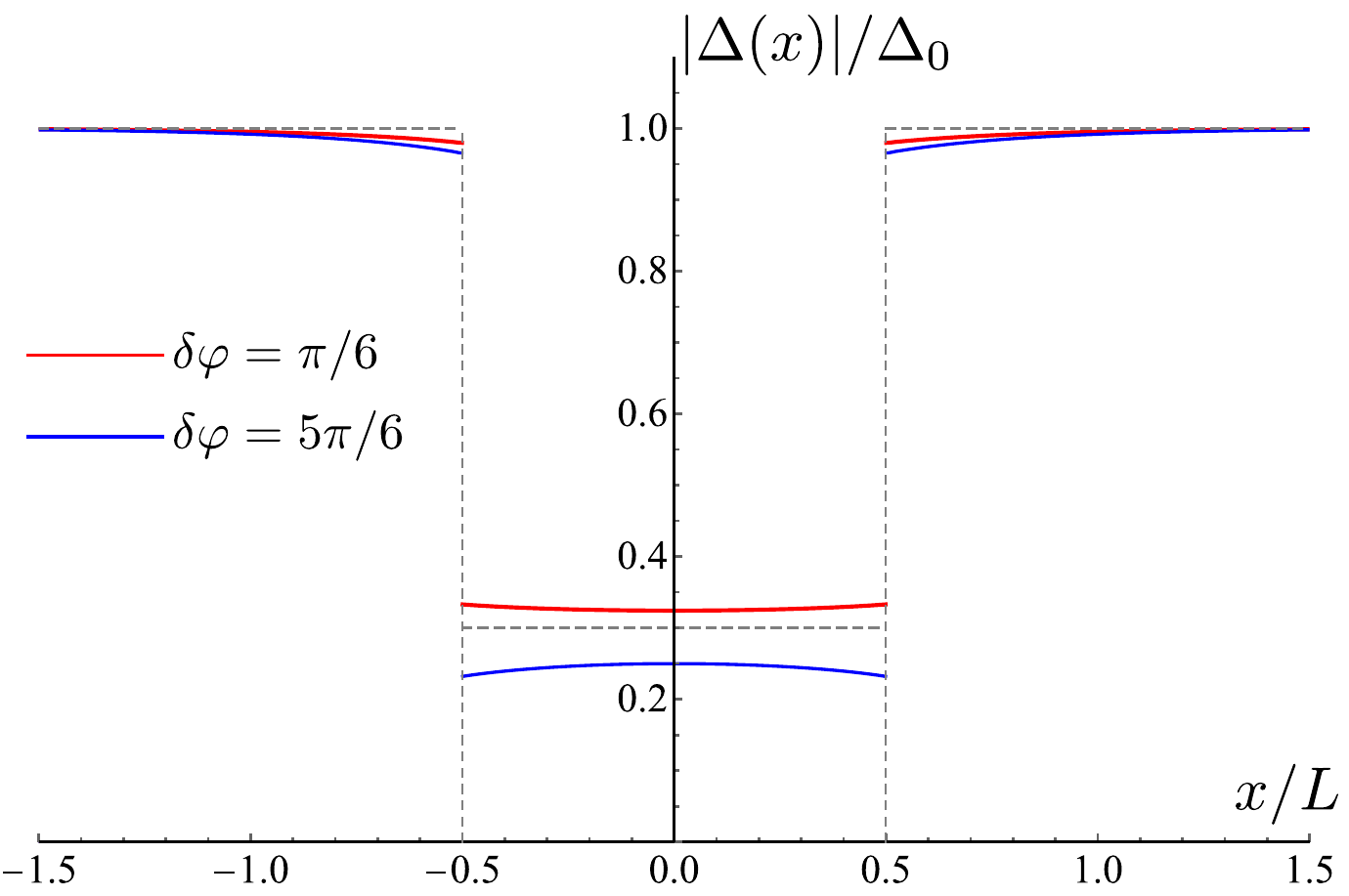}
    \caption{The absolute value of the OP in the S and S$^\prime$ regions.
    The OP contains the zero current value as well as the first correction in the $\alpha$ and $\alpha'$ parameters in accordance with Eqs.~\eqref{eq:resS} and \eqref{eq:resS'}.
    The corrections in the two regions are given by Eqs.~\eqref{eq:DeltaSimpleAns} and \eqref{eq:DeltaPrSimpleAns}, respectively for the two representative phase jumps at the right insulating barrier $\delta \varphi = \pi/6$ (red) and $\delta \varphi = 5 \pi/6$ (blue).
    The ratio of the zero current OPs (black dashed) is $\Delta^{\prime}_{0}/\Delta_{0}=0.3$, $T=0.18T_c$, $\xi=0.32L$, $\xi^{\prime}=0.58L$, $\alpha=0.032$, $\alpha^{\prime}=0.058$. 
    The OP is normalized to $\Delta_{0}$.}
    \label{fig:Delta1}
\end{figure}

\subsection{Order parameter absolute value}\label{sec:OPabs}

We start with the results for the absolute value of the OP. 
The details of the calculations are presented in Appendix~\ref{AppendixSRegionOrderParameter} for the S region and in the Appendix~\ref{AppendixS'RegionOrderParameter} for the S$^\prime$ region.
As explained above, it is enough to state the expressions for the OP in the S region for positive coordinate, $x> L/2$. In terms of the dimensionless coordinate $z=(x-L/2)/\xi >0$, 
\begin{align}
\label{eq:DeltaSimpleAns}
\frac{\Delta_{1}(z)}{\Delta_{0}}=& -2\left(\frac{\Delta_{0}}{\Delta_{0}^{\prime}}-\cos\delta\varphi\right)
\notag \\
& \times \intop_{-\infty}^{+\infty}\frac{dk}{2\pi}e^{ikz}\frac{L_{1,1}(k)-L_{3,1}(k)}{k^{2}L_{2,0}(k)+L_{3,0}(k)}\, , 
\end{align}
where for compactness we have introduced the notations:
\begin{subequations}
\label{eq:Ldefine}
\begin{align}
L_{n,m}(k) & = \frac{2\pi T}{\Delta_{0}}\sum_{\omega>0}\frac{\sin^{n}\theta_{0}\sin^{m}\theta_{0}^{\prime}}{k^{2}\sin\theta_{0}+1},\\
 L_{n,m}^{\prime}(k)& = \frac{2\pi T}{\Delta_{0}^{\prime}}\sum_{\omega>0}\frac{\sin^{n}\theta_{0}^{\prime}\sin^{m}\theta_{0}}{k^{2}\sin\theta_{0}^{\prime}+1}.
\end{align}
\end{subequations}
As we ignore the pair breaking, in Eq.~\eqref{eq:Ldefine} we have $\sin \theta_0 = \Delta_0/\sqrt{\Delta_0^2 + \omega_n^2}$, and $\sin \theta'_0 = \Delta'_0/\sqrt{\Delta_0^{\prime 2} + \omega_n^2}$, and the summation runs over the positive Matsubara frequencies $\omega_n=(2n+1)\pi T$. 

For the S$^\prime$ region we have in terms of the dimensionless coordinate $y=(x-L/2)/\xi'$ ($-L/\xi' < y < 0$),
\begin{align}\label{eq:DeltaPrSimpleAns}
\frac{\Delta_{1}^{\prime}(y)}{\Delta_{0}^{\prime}}& =-2\frac{\xi^{\prime}}{L}\left(\frac{\Delta_{0}^{\prime}}{\Delta_{0}}-\cos\delta\varphi\right)
\notag \\
\times & \sum_{n=-\infty}^{\infty}e^{ik_{n}y}\frac{L_{1,1}^{\prime}(k_{n})-L_{3,1}^{\prime}(k_{n})}{k_{n}^{2}L_{2,0}^{\prime}(k_{n})+L_{3,0}^{\prime}(k_{n})}\, ,
\end{align}
where $k_{n}=2\pi n/(L/\xi^{\prime})$.

\begin{figure}[t!]
    \includegraphics[width=0.45\textwidth]{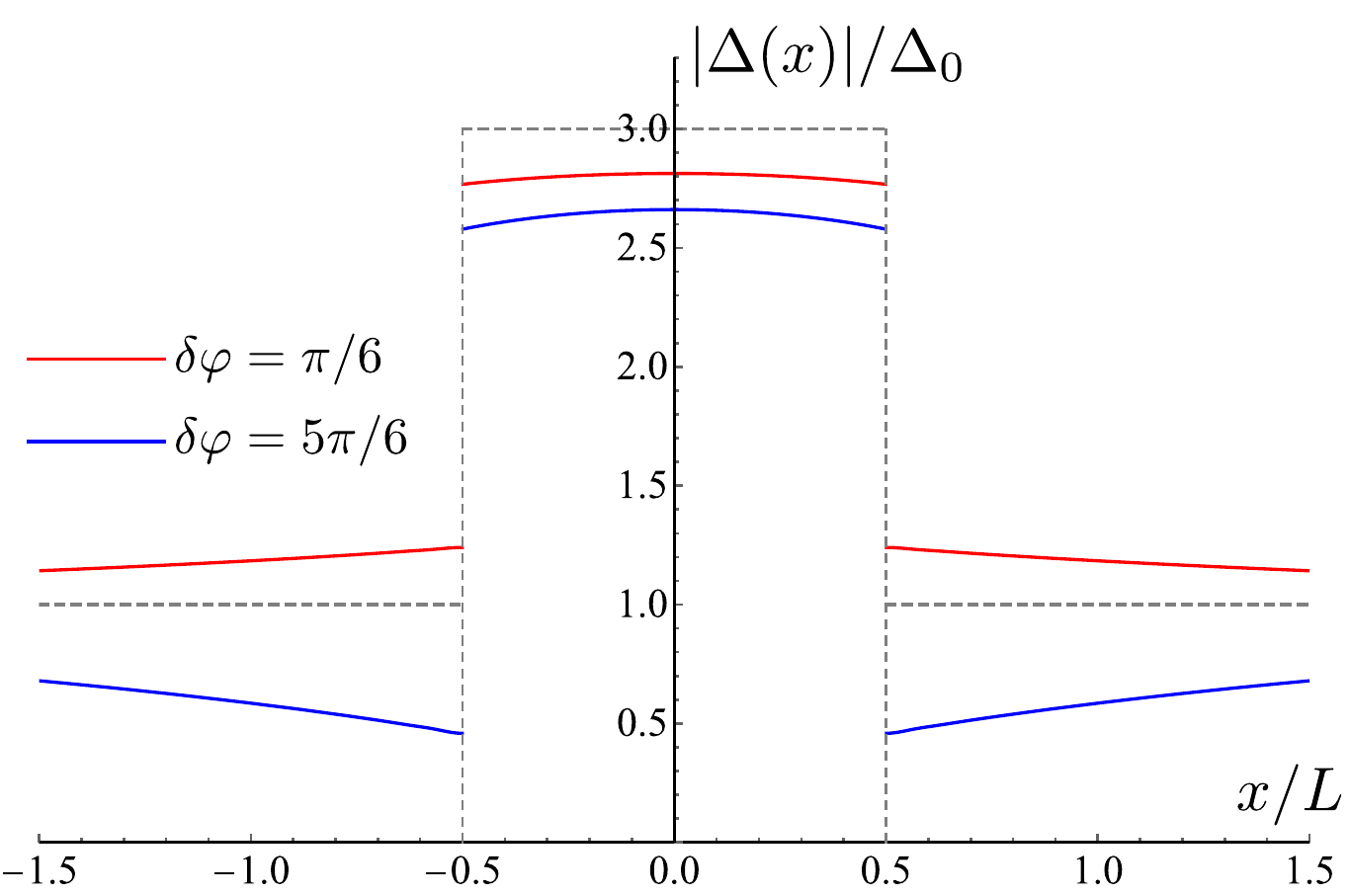}
    \caption{All the parameters are the same as in Fig.~\ref{fig:Delta1} except for the ratio, $\Delta^{\prime}_{0}/\Delta_{0}=3$.}
    \label{fig:Delta2}
\end{figure}

The expressions given by Eqs.~\eqref{eq:DeltaSimpleAns} and \eqref{eq:DeltaPrSimpleAns} are illustrated in Figs.~\ref{fig:Delta1} and \ref{fig:Delta2} for $\Delta_0 > \Delta'_0$ and $\Delta'_0 > \Delta_0$, respectively.
At the interface, for the phase jump $\delta \varphi \lesssim \pi /2$ the larger one of the two OPs decreases, and the smallest one increases. 
However, for $\delta \varphi \gtrsim \pi /2$ both OPs are suppressed.
Qualitatively, these results can be understood as different manifestations of the proximity effect.

To illustrate this point consider $\delta\varphi\approx0$. 
In this case, the OP can be chosen to be a positive real number on both sides of the interface.  
The proximity effect is the tendency of the OPs to approach each other.
This means that near the interface the larger (smaller) of the two OPs is suppressed (grows). 
When $\delta\varphi\approx\pi$, the OP remains a real number.
Yet, in contrast to the previous case $\delta\varphi\approx0$ it changes sign across the interfaces. 
As before, the OPs tend to equalize.
Now it implies that the positive (negative) OPs gets a negative (positive) correction at the interface, respectively.
Hence the absolute value of the OP is everywhere suppressed.

\subsection{Order parameter phase}\label{sec:OPphase}

\begin{figure}
    \includegraphics[width=0.45\textwidth]{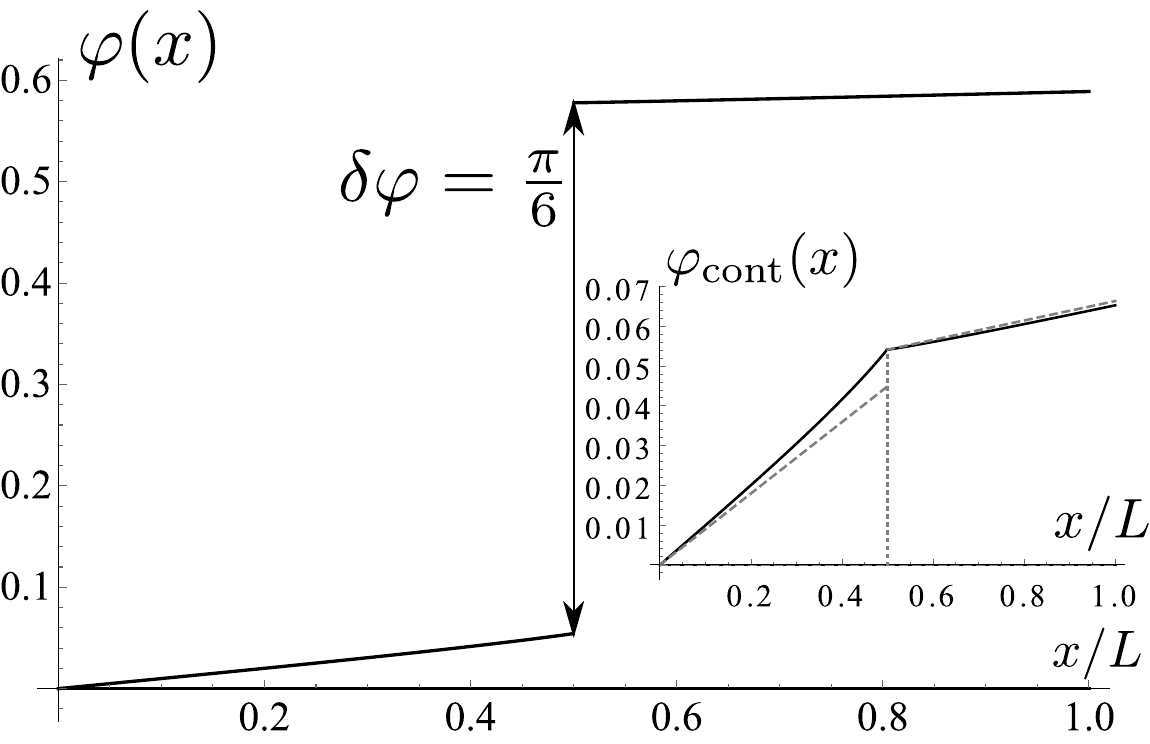}
    \caption{The spatial variation of the phase plotted for the same set of parameters as in the Fig.~\ref{fig:Delta1} with the the phase jump set to $\delta \varphi = \pi/6$. The absolute value of the OP is presented in Fig.~\ref{fig:Delta1} (red). The inset shows the spatial variation of the phase  with the discontinuity removed for clarity. The dashed lines represent the linear phase dependencies in the depth of the S$^{\prime}$ and S regions. For clarity, the lines are drawn so that inside the S$^{\prime}$ layer, the line turns to zero at $x=0$, and in the S layer, so that the line plot leaves the phase jump point without vertical shift.
    }
    \label{fig:phase1}
\end{figure}

We now turn to the discussion of the phase of the OP.
The details of this calculation are presented in App.~\ref{AppendixSRegionPhases}.
In the S region we have
\begin{equation}\label{eq:Varphi1SimpleAns}
\varphi_{1}(z)=2\sin\delta\varphi\intop_{\mathbb{-\infty}}^{+\infty}\frac{dk}{2\pi}\frac{1-\cos kz} {k^{2}}\frac{L_{1,1}(k)}{L_{2,0}(k)} + \varphi_{1}(+0).
\end{equation}
To find the constant $\varphi_{1}(+0)$ we use the fact that the jump of the phase at the boundary is fixed to $\delta \varphi$ and by definition has no corrections to all orders in $\alpha$ ($\alpha'$) which implies 
\begin{align}\label{eq:phase_cont}
   \alpha \varphi_{1}(+0) = \alpha' \varphi_1'(-0) \, .
\end{align}
The result for the phase in the S$^\prime$ region is derived in Appendix~\ref{AppendixS'RegionPhases} and reads as 
\begin{equation}
\varphi_{1}^{\prime}(y)=\sin\delta\varphi\frac{2\xi^{\prime}}{L}\sum_{n=-\infty}^{+\infty}\frac{\cos(k_{n+1/2}y)}{k_{n+1/2}^{2}}\frac{L_{1,1}^{\prime}(k_{n+1/2})}{L_{2,0}^{\prime}(k_{n+1/2})}\, .\label{eq:VarphiPrSimpleAns}
\end{equation}
Equations~\eqref{eq:phase_cont} and \eqref{eq:VarphiPrSimpleAns} specify the additive constant in Eq.~\eqref{eq:Varphi1SimpleAns}.
The expressions given in Eqs. \eqref{eq:Varphi1SimpleAns} and \eqref{eq:VarphiPrSimpleAns} are illustrated in Fig.~\ref{fig:phase1}.
These results along with Eqs.~\eqref{eq:DeltaSimpleAns} and \eqref{eq:DeltaPrSimpleAns} fully specify the self-consistent OP in the presence of a finite current to the first order in the considered perturbation theory.

\subsection{Josephson current}\label{sec:J}

As the current-carrying state is defined by the phase jump $\delta \varphi$ at the two interfaces, $x= \pm L/2$, 
$\varphi_1(\pm L/2) = \varphi'_1(\pm L/2) = 0$ by construction.
Although $\delta \varphi$ is a convenient parameter, it is not directly measured. For this reason below, at the end of this section, we reformulate the results in terms of the full phase change $\Delta \varphi$ including in addition to $2 \delta \varphi$ the phase accumulated in the S$^\prime$ region.
Throughout the section any possible pair breaking is ignored.

For the current we present the result as an expansion in the parameters $\alpha$ and $\alpha'$ similar to Eqs.~\eqref{eq:resS} and \eqref{eq:resS'} as follows: 
\begin{align}\label{eq:resJ}
    J(\delta \varphi) = J_1(\delta \varphi) + J_2^{(1)}(\delta \varphi)+ J_2^{(2)}(\delta \varphi)\, ,
\end{align}
where the subscript denotes the order in $\alpha$ ($\alpha'$).
To the second order we obtain the correction to the first Fourier harmonics in $\delta \varphi$ 
as well as the second Fourier harmonics denoted in Eq.~\eqref{eq:resJ} as $J_2^{(1)}$ and $J_2^{(2)}$, respectively.
The details of the derivation leading the three contributions are outlined in Appendix~\ref{AppendixCPR}.

The leading contribution to the current in Eq.~\eqref{eq:resJ}, 
\begin{equation}\label{eq:Current1Order}
J_{1}(\delta\varphi)=\sin\delta\varphi \frac{2\pi T}{eR_{\text{int}}}\sum_{\omega>0}
\frac{\Delta_0}{\sqrt{\Delta_0^2 + \omega_n^2}}\frac{\Delta'_0}{\sqrt{\Delta_0^{'2} + \omega_n^2}},
\end{equation}
has a form of Ambegaokar-Baratoff formula \cite{Ambegaokar1963} expressed solely via the resistance of a single barrier, $R_{\text{int}} = G^{-1}_{\mathrm{int}}$. 
Here and throughout the section we ignore the pair-breaking effects for clarity. 
The current phase relation, \eqref{eq:Current1Order} takes the form $J_{1}(\delta\varphi)= J_c \sin\delta\varphi$ defining the critical current $J_c$ which at $T=0$ reads as \cite{Abrikosov1988}
\begin{equation}
J_c=\frac{2\Delta_0\Delta'_0}{eR_{\text{int}}(\Delta_0+\Delta'_0)}K\left(\frac{|\Delta_0-\Delta'_0|}{\Delta_0+\Delta'_0}\right),
\end{equation}
where $K(k)=\int^{\pi/2}_{0}(1-k^2\sin^2\theta)^{-1/2}d\theta$ is the complete elliptical integral of the first kind. For a symmetric junction, $K(0)=\pi/2$, therefore, $J_c=\pi\Delta_0/2eR_{\text{int}}$ \cite{Ambegaokar1963}.  

The corrections to the OP and the phase to the first order in $\alpha$ ($\alpha'$) allow us to obtain the Josephson current to the second order in the same parameters. Specifically, we obtain the leading correction to the known expression  \eqref{eq:Current1Order}. This is achieved by computing the current at the interface directly from the boundary conditions and using the results listed in Secs.~\ref{sec:OPabs} and \ref{sec:OPphase}:
\begin{subequations}\label{eq:J2Ans}
    \begin{align}
    J_{2}^{(1)}(\delta\varphi)=&-\frac{\sin\delta\varphi}{eR_{\text{int}}}
    \Big[
    \alpha\frac{\Delta_{0}^{2}}{\Delta_{0}^{\prime}}\intop_{\mathbb{-\infty}}^{+\infty}\frac{dk}{\pi}S(k)
    \notag \\
    + & \alpha^{\prime}\frac{\Delta_{0}^{\prime2}}{\Delta_{0}}\frac{2\xi^{\prime}}{L}\sum_{n=-\infty}^{+\infty}S^{\prime}(k_{n})
    \Big],\label{eq:AppendixJ21Ans}\\
    J_{2}^{(2)} (\delta\varphi)&=\frac{\sin2\delta\varphi}{2eR_{\text{int}}}
    \Big\{ \alpha\Delta_{0}\intop_{\mathbb{-\infty}}^{+\infty}\frac{dk}{\pi}\left[S(k)+P(k)\right]
    \notag \\
    + & \alpha^{\prime}\Delta_{0}^{\prime}\frac{2\xi^{\prime}}{L}\sum_{n=-\infty}^{+\infty}\left[S^{\prime}(k_{n})+P^{\prime}(k_{n+1/2})\right]
    \Big\} \label{eq:AppendixJ22Ans}
\end{align}
\end{subequations}
Here, for brevity we have introduced the auxiliary functions
\begin{subequations}
\begin{align}
    &S(k)\equiv\frac{\left[L_{1,1}(k)-L_{3,1}(k)\right]^{2}}{k^{2}L_{2,0}(k)+L_{3,0}(k)}+L_{1,2}(k)-L_{3,2}(k),\label{eq:AppendixSDef}\\
    &P(k)\equiv L_{2,1}(k)\cdot\frac{L_{1,1}(k)}{L_{2,0}(k)}-L_{1,2}(k),\label{eq:AppendixPDef}\\
    &S^{\prime}(k)\equiv\frac{\left[L_{1,1}^{\prime}(k)-L_{3,1}^{\prime}(k)\right]^{2}}{k^{2}L_{2,0}^{\prime}(k)+L_{3,0}^{\prime}(k)}+L_{1,2}^{\prime}(k)-L_{3,2}^{\prime}(k),\label{eq:AppendixSPrDef}\\
    &P^{\prime}(k)\equiv L_{2,1}^{\prime}(k)\cdot\frac{L_{1,1}^{\prime}(k)}{L_{2,0}^{\prime}(k)}-L_{1,2}^{\prime}(k).\label{eq:AppendixPPrDef}
\end{align}
\end{subequations}
The correction to the first harmonic, Eq.~\eqref{eq:J2Ans}, has a negative sign with respect to the leading contribution, \eqref{eq:Current1Order} as detailed in App.~\ref{app:signJ12}.

\begin{figure}
\includegraphics[width=0.95\columnwidth]{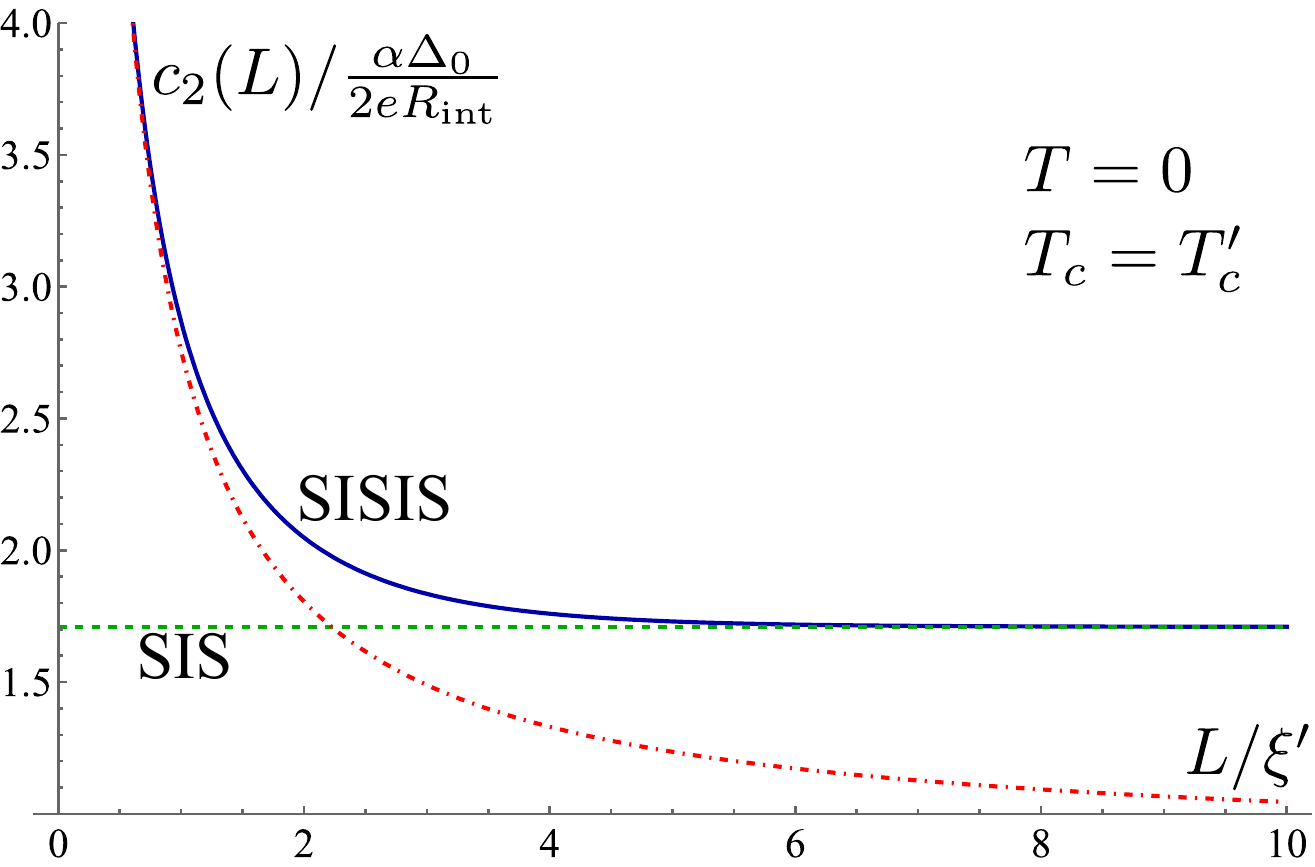}
\caption{This plot shows the amplitude, $c_2$, of the second harmonics $J^{(2)}_2(\delta\varphi)=c_2 \sin 2\delta\varphi$ [see Eq.~\eqref{eq:resJ}] as a function of the junction length $L$. 
For symmetric junction, $c_2$ is conveniently normalized to $\alpha \Delta_0/ 2 e R_{\mathrm{int}}$. 
The junction here is to be considered at $T=0$ and to be symmetric, $D=D'$, $\ell=\ell'$, $T_c=T^{\prime}_c$ (blue curve, SISIS). In the limit $L \gg \xi'$ the system passes to the symmetric SIS junction, the result for which was obtained earlier in \cite{Osin2021} and is shown here by a green dashed line.  
In the limit $L \rightarrow 0$ the increasing role of proximity effect determines the second harmonic amplitude and leads to nonperturbative effects \cite{AL2006,Whisler2018}. The red dot dashed line shows the asymptotics of the answer in the limit $L\ll\xi^\prime$, $c_2(L)=\alpha \Delta_0/ 2 e R_{\mathrm{int}}[0.86+(16+3\pi^2)\xi^\prime/24L]$, obtained from the formulas ~\eqref{eq:AppendixJ22Ans}, ~\eqref{eq:AppendixShortDeltaAns} and  ~\eqref{eq:AppendixShortVarphi1Ans}.}
\label{fig:c2}
\end{figure}

In Fig.~\ref{fig:c2} we plot the amplitude of the second harmonic as a function of the length, $L$ of the S$^\prime$ region.
As expected when the two superconductors are identical the results match the expression obtained previously for the SIS system in the limit $L \gg \xi'$.

\begin{figure}
    \centering
    \includegraphics[width=0.95\columnwidth]{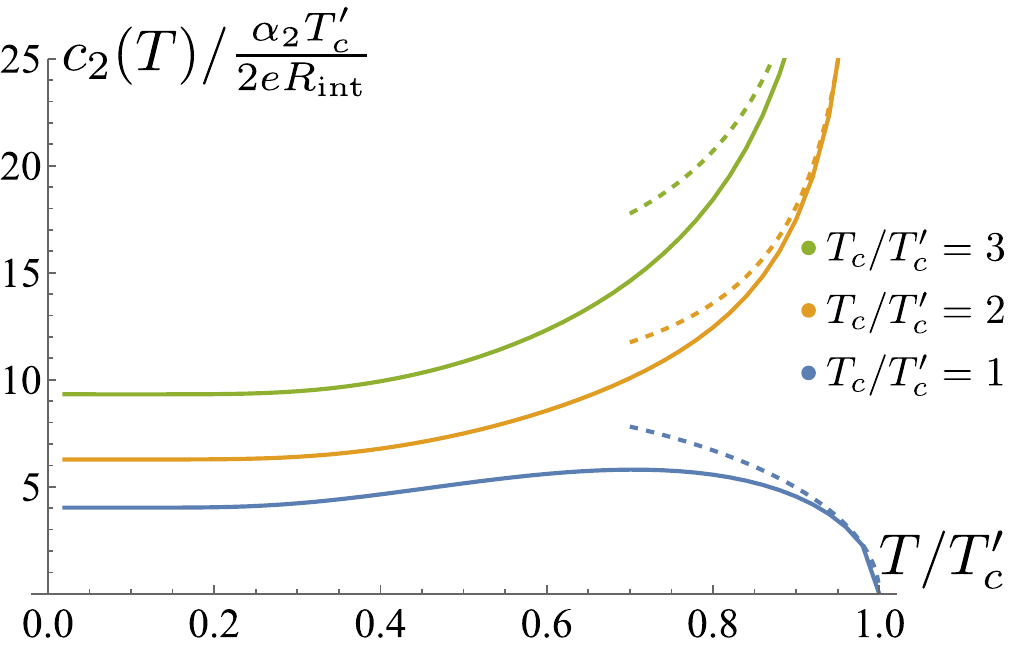}
    \caption{Temperature dependence of the second harmonics amplitude plotted for the different ratios between $T_c$ and $T^{\prime}_c$. The junction is considered in the regime $L\gg \xi'$, $D=D'$, $\sigma=\sigma'$, $\alpha_{2}=g_{\text{int}}\sqrt{D/2\pi T^{\prime}_c}/\sigma$. For $T_c \neq T^{\prime}_c$ the second harmonic amplitude grows in a monotonic way which is due to the increasing importance of the proximity effect in S$^\prime$ region. 
    At some temperatures sufficiently close to  $T^{\prime}_c$ the perturbation theory breaks down. 
    In contrast, for symmetric junction, $T_c = T^{\prime}_c$ after an initial growth the second harmonic decreases in agreement with previous works (see Refs. \cite{Osin2021,Kupriyanov1982}). The dashed curves are asymptotic curves of behavior $c_2(T)$ obtained in the formulas ~\eqref{eq:AppendixSPrGLShortJ22Ans} (blue), ~\eqref{eq:AppendixSPrGLLongJ22Ans} (yellow and green). To recalculate in terms of $\delta\varphi$ instead of $\Delta\varphi$, the $\alpha^{\prime}_L$ term must be excluded from the above formulas.}
    \label{fig:c2T}
\end{figure}

As we can see, the second harmonic grows substantially for relatively thin S$^\prime$ region, $L \lesssim \xi'$. The perturbation theory we have developed in this work fails to properly describe length dependence in this limit. The reason for this is the strong proximity effect in this limit \cite{AL2006,Whisler2018}. The proximity effect, on the other hand, is most pronounced when $T_c^\prime < T_c$ and $T$ approaches $T_c^\prime$. Therefore, in this regime one expects the break down of the perturbation theory. This is demonstrated in Fig.~\ref{fig:c2T}.

The current in Eq.~\eqref{eq:resJ} is expressed as a function of the discontinuity $\delta \varphi$ that the OP phase, $\varphi$, experiences at each of the two interfaces.
This parametrization of the current carrying state is natural in view of the perturbation theory we applied to solve the problem.

Experimentally, however, the current is measured in terms of the total phase change $\Delta \varphi$ that includes the two jumps $\delta \varphi$  at the barriers as well as the phase $\varphi_{\text{cur}}= \varphi(L/2-0)-\varphi(-L/2+0)$ accumulated in the region $S^\prime$ sandwiched between the barriers.
By definition, the two phases $\delta \varphi$ and $\Delta \varphi$ are related as
\begin{equation}
\Delta\varphi=2\delta\varphi+\varphi_{\text{cur}}+2\pi n,\label{eq:AppendixDeltaVarphiDef}
\end{equation}
where $n \in \mathbb{Z}$ expresses the general property of the phase of being defined modulo $2 \pi$.

The goal is to write the current in the form
\begin{equation}
J(\Delta\varphi)=J_{1}(\Delta\varphi)+J_{2}^{(1)}(\Delta\varphi)+J_{2}^{(2)}(\Delta\varphi),\label{eq:AppendixCPRDeltaVarphiAns}
\end{equation}
instead of \eqref{eq:resJ}.
The separate contributions in Eq.~\eqref{eq:AppendixCPRDeltaVarphiAns} are obtained in Appendix~\ref{AppendixCPRDeltaVarphi}:
\begin{subequations}\label{eq:JDelta12}
    \begin{align}
        J_{1}(\Delta\varphi)=(-1)^{n}\sin\left(\frac{\Delta\varphi}{2}\right)\frac{\Delta_{0}}{eR_{\text{int}}}L_{1,1}(0),\label{eq:AppendixJ1DeltaVarphiAns}
 \end{align}
\begin{align}
        &J_{2}^{(1)}(\Delta\varphi)=-\frac{(-1)^{n}\sin\left(\Delta\varphi/2\right)}{eR_{\text{int}}}
        \notag \\
        &\times \Bigg[\alpha\frac{\Delta_{0}^{2}}{\Delta_{0}^{\prime}}\intop_{\mathbb{-\infty}}^{\infty}\frac{dk}{\pi}S(k) + \alpha^{\prime}\frac{\Delta_{0}^{\prime2}}{\Delta_{0}}\frac{2\xi^{\prime}}{L}\sum_{n=-\infty}^{\infty}S^{\prime}(k_{n})\Bigg],\label{eq:AppendixJ21DeltaVarphiAns}
\end{align}
\begin{align}
        J_{2}^{(2)}(\Delta\varphi)=&
        \frac{\sin\Delta\varphi}{2eR_{\text{int}}}
        \Bigg\{ \alpha\Delta_{0}\intop_{\mathbb{-\infty}}^{\infty}\frac{dk}{\pi}\left[S(k)+P(k)\right]
        \notag \\
        + & 
        \alpha^{\prime}\Delta_{0}^{\prime}\frac{2\xi^{\prime}}{L}\sum_{n=-\infty}^{\infty}\left[S^{\prime}(k_{n})+P^{\prime}(k_{n+1/2})\right]
        \notag \\
        - & \frac{\alpha_{L}^{\prime}}{2}\Delta_{0}L_{1,1}(0)R(L/\xi^{\prime})\Bigg\}.
        \label{eq:AppendixJ2DeltaVarphiAns}
    \end{align}
\end{subequations}
The current-phase relationship \eqref{eq:AppendixCPRDeltaVarphiAns} is $2\pi$ periodic.
This is achieved by a suitable choice of a sign in Eqs.~\eqref{eq:AppendixJ1DeltaVarphiAns},
\eqref{eq:AppendixJ21DeltaVarphiAns}, i.e., choosing the number $n$.
The number $n$ is determined by the interval in which the value $\Delta\varphi$
lies $\Delta\varphi\in\left(\pi(2n-1);\pi(2n+1)\right)$. Thus, the factor $(-1)^{n}$ in the formulas \eqref{eq:AppendixJ1DeltaVarphiAns} and  \eqref{eq:AppendixJ21DeltaVarphiAns} can be replaced by $\sgn\left[\cos\left(\Delta\varphi/2\right)\right]$. The dimensionless function $R(t)$ is defined by Eq. \eqref{eq:AppendixRDef}. 

\begin{figure}[t!]
\includegraphics[width=0.95\columnwidth]{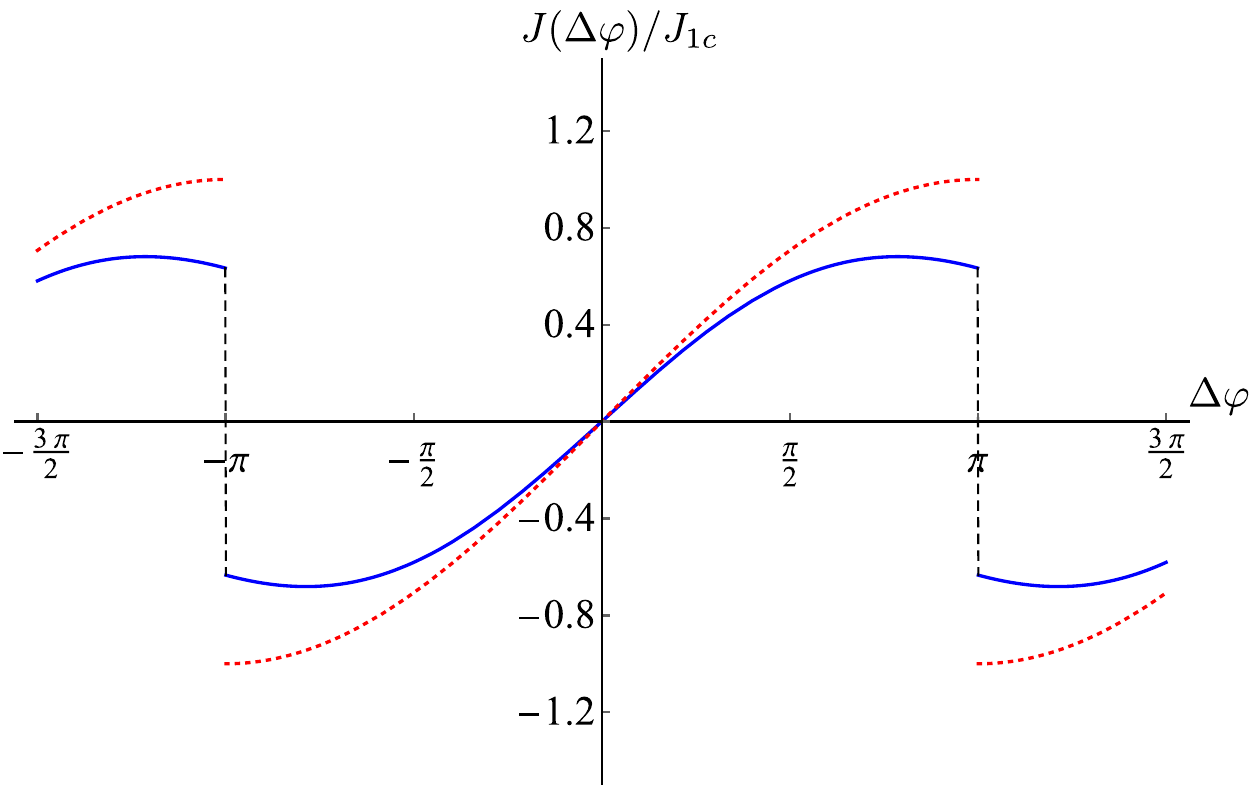}
\caption{The current-phase relation plotted for a symmetric junction at zero temperature shown as solid (blue) curve. The interface conductances were chosen to be such that $\alpha=\alpha'=0.2$ and the length of the middle layer is set equal to the coherence length $L=\xi'$. The overall scale of the current is normalized in units of $J_{1c}=\pi\Delta_0/2eR_{\text{int}}$. For comparison we also presented CPR without the correction terms, namely $J(\Delta\varphi)=J_{1c}\sin(\Delta\varphi/2)\sgn[\cos(\Delta\varphi/2)]$ shown as dashed (red) curve.}
\label{fig:CPR}
\end{figure}

In Fig. \ref{fig:CPR} we plot the current-phase relation of SIS$^\prime$IS junction with an account of the correction terms. Finally, to take into account the anomaly in the current-phase relation, one has to replace the $J(\Delta \varphi)$ as given by Eqs.~\eqref{eq:AppendixCPRDeltaVarphiAns} and \eqref{eq:JDelta12} by $J(\Delta\varphi-2eA_{\text{eff}}L)$ in accordance with Eq.~\eqref{eq:resS}.

\section{Summary of the solution method}\label{sec:UsadelBC}

In this section, we formulate the Usadel equations~\cite{Usadel1970,Svidzinsky1982} for S and S$^\prime$ regions separated by an insulating interfaces. We consider a particular realization of the junction such that the spin-orbit coupling is absent (present) in S (S$^\prime$) regions, respectively. We then supplement these equations with the proper boundary conditions in order to match the solutions across the interfaces that respect continuity of current.

\subsection{Usadel equations in S and S$^\prime$ regions}

We consider an infinitely long planar strip subjected to the in-plane (Zeeman) magnetic field. In this limit the S regions are infinitely long and straight. 
As in a standard situation, a finite vector potential $\mathbf{A}$ exists in both S and S$^\prime$ regions. 
We choose the gauge in such a way that the vector potential associated with the in-plane magnetic field vanishes within the junction.
This reflects the fact that the in-plane magnetic field does not couple to the orbital degrees of freedom in the considered geometry.

The Usadel equations \cite{Usadel1970} are obeyed by the quasi-classical Green's function, $\check{g}(\mathbf{r},\omega_n)$ which in the case of spin-singlet pairing is a $2 \times 2$ matrix in the particle-hole (Nambu) space.
The Usadel equations are discussed separately for S and S$^\prime$ regions below.

\subsubsection{S regions}

We start with the S region, where the Usadel equations have a standard form:
\begin{align}\label{US_13}
     i  D \hat{\boldsymbol{\nabla}}(\check{g}  \hat{\boldsymbol{\nabla}}  \check{g}) + [\check{H},\check{g}]  = 0\, , 
\end{align}
where the covariant derivative includes magnetic field  
\begin{align}\label{US_15}
    \hat{\boldsymbol{\nabla}} \check{g} =  \boldsymbol{\nabla} \check{g}- i e\mathbf{A} [\check{\tau}_z, \check{g}]\, ,\quad 
\check{H} = \begin{pmatrix} -i \omega_n & -\Delta \\ \Delta^* &  i \omega_n \end{pmatrix}\, .
\end{align}
Here $[\hat{a},\hat{b}]$ stands for the commutation relation, and $\omega_n=\pi T\left(2n+1\right),\ n\in\mathbb{Z}$ is the Matsubara frequency, and $\check{\tau}_{i}$ are unit ($i=0$), and Pauli matrices for $i =x,y,z$ operating in the Nambu space. 

The quasiclassical Green's function is normalized $\check{g}^2 = \tau_0$. In order to resolve this constraint, we employ the trigonometric parametrization \cite{Belzig1999}:
\begin{align}\label{US_17}
    \check{g}(\mathbf{r},\omega_n) = \begin{pmatrix} \cos \theta & - i \sin \theta e^{ i \chi} \\ i \sin \theta e^{- i \chi}& - \cos \theta  \end{pmatrix}\, ,
\end{align}
where $\theta(\mathbf{r},\omega_n)$ is the superconducting angle, $\chi(\mathbf{r},\omega_n)$ is the phase of the anomalous Green's function.
For a fixed Matsubara frequency $\omega_n$ the matrix, $\check{g}(\omega_n)$ is Hermitian. This implies that the independent parameters $\theta$ and $\chi$ are real valued.

The gauge-invariant Usadel equations in the S regions read as
\begin{subequations}\label{US_21}
\begin{align}\label{eq:Diffusion1D}
    \frac{D}{2}\frac{d^{2}\theta}{dx^{2}}
    +&\left|\Delta\right|\cos\left(\chi-\varphi\right)\cos\theta-\omega_n\sin\theta
    \notag \\
    & -\frac{D}{2}\sin\theta\cos\theta\left(\frac{d\chi}{dx}\right)^{2}=0\, ,
\end{align}
\begin{align}\label{eq:Continuity1D}
    \frac{D}{2}\frac{d}{dx}
    \left(\frac{d\chi}{dx}\sin^{2}\theta\right)
=\left|\Delta\right|\sin\left(\chi-\varphi\right)\sin\theta\, .
\end{align}
\end{subequations}

In Eq.~\eqref{US_21} the OP has to be determined from the self-consistency condition, 
\begin{align}\label{eq:SelfConsistency1D}
    \left|\Delta\right|=\pi\lambda T\sum_{\left|\omega\right|<\omega_{D}}e^{i\left(\chi-\varphi\right)}\sin\theta\, ,
\end{align}
where $\lambda$ is the dimensionless coupling constant, and $\omega_D$ is the Debye frequency. The supercurrent in the S region takes the form
\begin{align}\label{eq:Current1D}
    J=2\pi\nu_0 \mathcal{W}DTe\sum_{|\omega|<\omega_{D}}\sin^{2}\theta\frac{d\chi}{dx}\, .
\end{align}
The charge conservation, $d J/d x =0$ follows from Eq.~\eqref{eq:Continuity1D} and the self-consistency condition \eqref{eq:SelfConsistency1D}.

Far from the junction the superconductor in the S region has to approach the homogeneous current state in the infinite superconductor. 
As such a state provides the asymptotic to our solutions, we state the simplest solutions of the Usadel equations \eqref{US_21} describing it.

For a homogeneous current, the solution is achieved for spatially constant $d \chi/d x$ and $\theta$ implying $\chi = \varphi$ according to Eq.~\eqref{eq:Current1D}. 
The last term of Eq.~\eqref{eq:Diffusion1D} quadratic in phase gradient term describes the pair-breaking effect of the impurities in the presence of current.
It is responsible for the nonlinear Meissner effect \cite{Sauls2022}. 
This term is proportional to $\propto\alpha^2$ and therefore does not affect the solutions ~\eqref{eq:resS} and \eqref{eq:resS'} valid to the first order in $\alpha$.
In result, in this work we omit this term.

To the linear order in the phase gradients, the BCS solutions apply, namely:
\begin{equation}
\cos \theta = |\omega_n|/\sqrt{\Delta_0^2 + \omega_n^2},\quad  
\sin \theta = \Delta_0/\sqrt{\Delta_0^2 + \omega_n^2}.
\end{equation}
The current then reads from Eq.~\eqref{eq:Current1D},
\begin{align}\label{eq:J_hom1}
    J=\frac{\sigma}{e}\mathcal{W} \frac{d \varphi}{d x} \pi T \sum_{|\omega|<\omega_{D}}\frac{\Delta_0^2}{\omega_n^2 + \Delta_0^2}\equiv\frac{\mathcal{W}}{4\pi\lambda^2}\frac{1}{2e}\frac{d\varphi}{dx}.
\end{align}
Equation \eqref{eq:J_hom1} yields the standard result for the London penetration depth of a disordered superconductor, 
$\lambda^{-2}  = 4 \pi^2  \sigma  |\Delta_0| \tanh (|\Delta_0|/2 T)$ (see ~\cite{Svidzinsky1982}, Sec. 22, ~\cite{TinkhamBook}, Sec. 2.1\footnote{By adding the Maxwell's equation, $\nabla\times\mathbf{H}=4\pi\mathbf{j}$, to the Eq.~\eqref{eq:J_hom1},  and the connection between the gauge invariant phase gradient and the magnetic field $\nabla\times(\nabla\varphi-2e\mathbf{A})=-2e\mathbf{H}$, the penetration length can be obtained from the equation for the current $\nabla^{2}\mathbf{j}=\mathbf{j}/\lambda^2$. Here $\mathbf{j}$ is the current density.}).

\subsubsection{S$^\prime$ region}

The Usadel equations for the S$^\prime$ region have been derived in Ref.~\cite{Houzet2015}. In our parametrization defined by Eq. \eqref{US_17} they take the form
\begin{subequations}\label{US_31}
    \begin{align}\label{eq:Diffusion1D-1}
        & \frac{D^{\prime}}{2}\frac{d^{2}\theta^{\prime}}{dx^{2}} +\left|\Delta^{\prime}\right|\cos\left(\chi^{\prime}-\varphi^{\prime}\right)\cos\theta^{\prime}-\omega\sin\theta^{\prime}
        \notag \\
        & -\sin\theta^{\prime}\cos\theta^{\prime}\left[\frac{D^{\prime}}{2}\left(\frac{d\chi^{\prime}}{dx}-2eA_{\text{eff}}\right)^{2}+\Gamma^{\prime}\right]=0,
    \end{align}
\begin{align}\label{eq:Continuity1D-1}
    \frac{D^{\prime}}{2} 
    \frac{d}{dx}&\left[\left(\frac{d\chi^{\prime}}{dx}-2eA_{\text{eff}}\right)\sin^{2}\theta^{\prime}\right]
    \notag \\
    &=\left|\Delta^{\prime}\right|\sin\left(\chi^{\prime}-\varphi^{\prime}\right)\sin\theta^{\prime} \, .
\end{align}
\end{subequations}
The pair breaking parameter, $\Gamma^{\prime}$ results from the combination of the Zeeman field and the spin-orbit interaction, and for the Rashba-Zeeman system is given explicitly in Ref.~\cite{Houzet2015}. 
This term is present even at zero current and we keep it in Eq.~\eqref{US_31} for generality.
The finite $\Gamma'$ does not change our results qualitatively and we present the final expressions in the limit $\Gamma'=0$ for simplicity.

The self-consistency condition has the same form as 
Eq.~\eqref{eq:SelfConsistency1D},
\begin{align}\label{eq:SelfConsistency1D-1}    \left|\Delta^{\prime}\right|=\pi\lambda^{\prime}T\sum_{\left|\omega\right|<\omega_{D}^{\prime}}e^{i\left(\chi^{\prime}-\varphi^{\prime}\right)}\sin\theta^{\prime}\, ,
\end{align}
where in general the BCS coupling constants are different in the two regions, $\lambda \neq \lambda'$.
The expression for the current parallels Eq.~\eqref{eq:Current1D}, except for the additional effective vector potential  $A_{\mathrm{eff}}$: 
\begin{align}\label{eq:Current1D-1}
J=2\pi\nu_{0}^{\prime}\mathcal{W}D^{\prime}Te\sum_{|\omega|<\omega_{D}}\sin^{2}\theta^{\text{\ensuremath{\prime}}}\left(\frac{d\chi^{\prime}}{dx}-2eA_{\text{eff}}\right)\, .
\end{align}

\subsection{Boundary conditions}

In the limit of low transparency of the interfaces, one can use the 
Kupriyanov-Lukichev boundary conditions \cite{KL1987}.
In this case the conductance of the barrier per unit width, $g_{\text{int}} = G_{\text{int}}/\mathcal{W}$ is the only input parameter characterizing the barrier.
The reason for such a universality is the isotropization of the electron motion within the mean free path distance from the barrier.

Thanks to the symmetry of the system it is enough to focus on the $x=+L/2$ boundary.
In terms of the parametrization \eqref{US_17} the boundary conditions read as
\begin{subequations}\label{eq:KL}
    \begin{align}\label{eq:KLL}
        \sigma^{\prime}\frac{d\theta^{\prime}}{dx}\!=\!
        g_{\text{int}}\left[\cos\left(\chi-\chi^{\prime}\right)\cos\theta^{\prime}\sin\theta-\sin\theta^{\prime}\cos\theta\right],
    \end{align}
    \begin{align}\label{eq:KLR}
        \sigma\frac{d\theta}{dx}=g_{\text{int}}\left[\cos\theta^{\prime}\sin\theta-\cos\left(\chi-\chi^{\prime}\right)\sin\theta^{\prime}\cos\theta\right]\,,
    \end{align}
    \begin{align}\label{eq:KLCC}
     \frac{\sigma^{\prime}}{g_{\text{int}}}
     &
     \sin^{2}\theta^{\prime}
     \left(\frac{d\chi^{\prime}}{dx}-2eA_{\text{eff}}\right)=\frac{\sigma}{g_{\text{int}}}\sin^{2}\theta\frac{d\chi}{dx}
     \notag \\
     &=\sin\left(\chi-\chi^{\prime}\right)\sin\theta^{\prime}\sin\theta\, .
    \end{align}
\end{subequations}

Here $\theta^{\prime}(x,\omega)$ and $\chi^{\prime}(x,\omega)$ are
taken from the left of the boundary $x=L/2-0$ and $\theta(x,\omega)$,
$\chi(x,\omega)$ are taken to the right $x=L/2+0$, respectively. 

In addition to the boundary conditions at the interface, we must add
boundary conditions at infinity corresponding to the current-carrying
state of the system
\begin{gather}
\theta(x\to\infty,\omega)=\text{const},\label{eq:AppenixInftyTheta}\\
\frac{d\chi(x\to\infty,\omega)}{dx}=\frac{d\varphi(x\to\infty)}{dx}=\text{const}.\label{eq:AppenixInftyChi}
\end{gather}

Taking into account the boundary condition ~\eqref{eq:KLCC},
the expression for the current, Eqs.~\eqref{eq:Current1D} and~\eqref{eq:Current1D-1}, takes the form
\begin{equation}
J=\frac{\pi T}{eR_{\text{int}}}\sum_{|\omega|<\omega_{D}}\left[\sin\left(\chi-\chi^{\prime}\right)\sin\theta^{\prime}\sin\theta\right]_{x=L/2}.
\label{eq:AppendixCurrent}
\end{equation}

\subsection{Anomalous phase $\varphi_0$}

A finite anomalous phase shift,  $\varphi_0$ is a direct consequence of the finite $A_{\mathrm{eff}}$ in the S$^\prime$ region. The magnetic vector potential, $\mathbf{A}$ is pure gauge and is absorbed into a gauge-invariant phase controlling the current. Clearly, therefore, $\mathbf{A}$ cannot in principle give rise to a nonzero $\varphi_0$. This should be contrasted with the $A_{\mathrm{eff}}$ that is finite only within the S$^\prime$ region.
In this case gauging out $A_{\mathrm{eff}}$ gives rise to a finite phase across the junction.

To eliminate $A_{\mathrm{eff}}$ from Usadel equations ~\eqref{US_31} and from the boundary conditions \eqref{eq:KL} we perform the following transformation to the new variables denoted by the bars
\begin{align}\label{gauge_S'}
    \bar{\chi}'(x) = \chi'(x) + 2 e A_{\mathrm{eff}} x, 
    \notag \\
    \bar{\varphi}'(x) = \varphi'(x) + 2 e A_{\mathrm{eff}} x 
\end{align}
in the S$^\prime$ region.
To ensure the form of the boundary conditions does not change the transformation in Eq. \eqref{gauge_S'} should be accompanied by the corresponding transformation in the S region:
\begin{align}\label{gauge_S}
    \bar{\chi}(x) & = \begin{cases}
    \chi(x)  + 2 e A_{\mathrm{eff}} (L/2), & x>L/2 \\
    \chi(x)  - 2 e A_{\mathrm{eff}} (L/2), & x< - L/2
    \end{cases}
    \notag \\
    \bar{\varphi}(x) & = \begin{cases}
    \varphi(x)  + 2 e A_{\mathrm{eff}} (L/2), & x>L/2 \\
    \varphi(x)  - 2 e A_{\mathrm{eff}} (L/2), & x< - L/2.
    \end{cases}
\end{align}
In terms of the new, barred variables, the Usadel equations as well as the boundary conditions are the same as before the transformation with $A_{\mathrm{eff}}$ set to zero.

We make two observations at this point. First, transformations defined by Eqs. \eqref{gauge_S'} and \eqref{gauge_S} do not rely on a particular way of solving the Usadel equations and accompanying boundary conditions. The elimination of $A_{\mathrm{eff}}$ is entirely general. For instance, it holds to all orders in the perturbation theory developed in this work. Second, after the effective potential has been eliminated the phase of the current-phase relation shifts by a finite amount. In fact, in view of Eqs.~\eqref{gauge_S} and \eqref{gauge_S'} we have  
\begin{align}\label{shift}
\bar{\varphi}(L/2) -\bar{\varphi}(-L/2) = \varphi(L/2) -\varphi(-L/2) + \varphi_0\, ,
\end{align}
with the universal anomalous phase shift, $\varphi_0 =2 e A_{\mathrm{eff}} L$.
For instance, the current is zero for $\bar{\varphi}(L/2) =\bar{\varphi}(-L/2)$.
Yet, this according to Eq. \eqref{shift} requires, $\varphi(L/2)= \varphi(-L/2) - \varphi_0 $. Stated differently, at zero phase difference, $\varphi(L/2) -\varphi(-L/2)=0$ the current is finite, since
$\bar{\varphi}(L/2) \neq \bar{\varphi}(-L/2)$.

Having eliminated $A_{\mathrm{eff}}$ we remove the bars from all the variables, and work with unbarred notations throughout the paper.
Hence, we solve the problem without $A_{\mathrm{eff}}$ with the understanding that its effect amounts to the simple phase shift, Eq.~\eqref{shift}.

\section{Conclusions}

In this paper, we have studied the current-phase relation of the quasi-2D SIS$^\prime$IS Josephson junctions in the diffusive limit.
We have considered the spin-orbit-coupled S$^\prime$ region placed in a parallel magnetic field. To compute the current we have employed the perturbation theory in the small ratio of the barrier conductance to the conductance of the normal-state material on a coherence length. The effective potential generated by the Lifshitz invariant can be eliminated to all orders in the perturbation theory, leading to the simple phase shift of the current-phase relation.
Hence, we conclude that in the dirty limit, the only effect of the Lifshitz invariant is the phase shift of the current-phase relation. 

To the first order in the above-mentioned perturbation theory, the current is given by the Ambegaokar-Baratoff formula. 
In this limit, the current is determined by the current-phase relation of the SIS$^\prime$ junction and is not sensitive to the spatial extent of the S$^\prime$ region. Besides, it is not sensitive to the proximity effect.

In contrast, to the second order in the perturbation theory, the current-phase relation: (i) acquires a finite second Josephson harmonics, (ii) is sensitive to the length of the S$^\prime$ region, and (iii) is affected by the proximity effect.
When the length of the S$^\prime$ region exceeds the coherence length, the properties of the SIS$^\prime$IS region are determined by the individual SIS$^\prime$ junctions. When S and S$^\prime$ have vastly different critical temperatures the solution is limited to the long junctions where the otherwise strong proximity effect is reduced.

\begin{figure}[t!]
    \includegraphics[width=0.95\columnwidth]{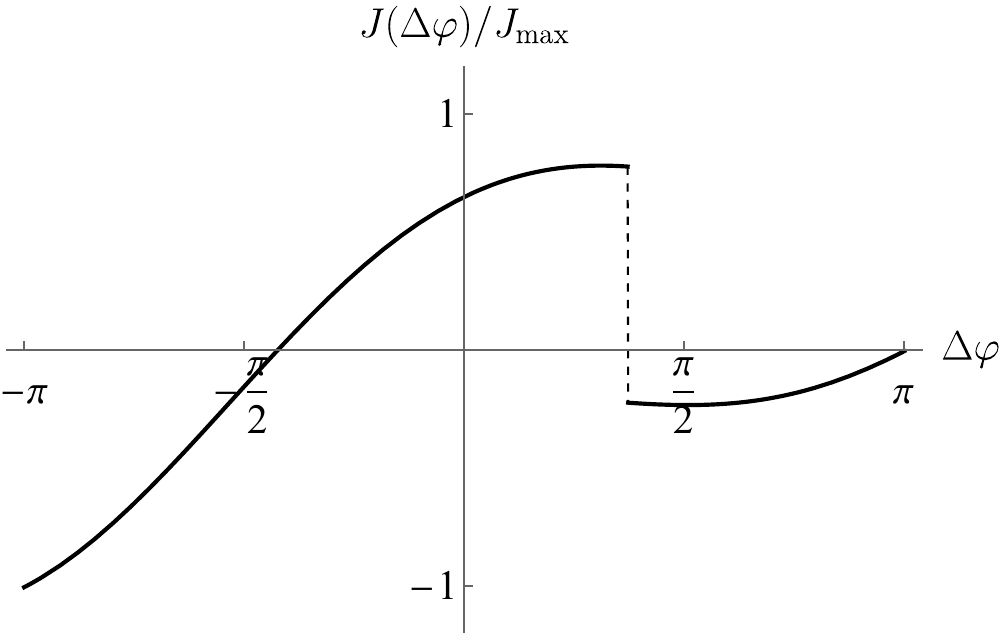}
    \caption{
    Finite SDE is obtained for a two-band superconductor SIS'IS junction. 
    The SDE arises as the current-phase relation of the two superconductors contains the second harmonics, and the OP in the two bands experience different $\varphi_0$ phase shifts.
     }
    \label{fig:SDE}
\end{figure}

In the simplest case of the single-band superconductors the critical current is symmetric with respect to the current direction (see Fig.~\ref{fig:CPR}). 
In contrast to the ballistic junction the SDE is absent in this limit.
The physical reason for this is that the multiple scattering off the disorder effectively replaces a large number of ballistic channels by a single collective diffusive channel.
This implies that the current-phase relation is phase shifted as a whole.

The SDE is finite if the superconductors under consideration contain more than one band \cite{Yerin-LTP17}.
The current-phase relations in both bands contain the second harmonics.
There are two conditions for the SDE in this case.
The ratio of the second to the first Josephson harmonics has to be different for the two bands.
And, in addition, the phase shifts for the two bands must be distinct.
Clearly, both conditions are generically satisfied. 
We illustrate this point in Fig.~\ref{fig:SDE}.

The considered mechanism of SDE is somewhat similar to the one proposed for the ballistic junctions.
It is, however, more universal in a sense that instead of infinitely many ballistic channels only two effective diffusive channels contribute. 
Experimentally, the second to the first harmonics ratio is controlled by the barrier thickness and temperature.
The difference in the two phase shifts is defined by the applied field as well as the length of the S$^\prime$ region. 
While the disordered single-band SIS$^\prime$IS structures have a reciprocal critical current,
the multiband SIS$^\prime$IS structure exhibits a finite SDE.
In the latter case the SDE can be experimentally controlled.

\acknowledgements

We thank I. Mazin for the useful comments. We acknowledge financial support from the Israel Science Foundation, Grant No. 2665/20 (A.O. and M.K.). This work at UW-Madison was financially supported by the National Science Foundation, Quantum Leap Challenge Institute for Hybrid Quantum Architectures and Networks Grant No. OMA-2016136. A.L. gratefully acknowledges H.I. Romnes Faculty Fellowship provided by the University of Wisconsin-Madison Office of the Vice Chancellor for Research and Graduate Education with funding from the Wisconsin Alumni Research Foundation.


\appendix

\begin{widetext}

\section{Perturbation theory}\label{AppendixPerturbation theory}

In this appendix we follow the general ideas of the method outlined
in Ref.~\cite{Osin2021}. 
In terms of the dimensionless variables $z\equiv(x-L/2)/\xi$
and $y=(x-L/2)/\xi^{\prime}$, the boundary conditions~\eqref{eq:KL} take the form ~\eqref{eq:KLL}--\eqref{eq:KLCC} as
\begin{gather}
\frac{d\theta^{\prime}(y=-0)}{dy}=\alpha^{\prime}\left[\cos\left(\chi-\chi^{\prime}\right)\cos\theta^{\prime}\sin\theta-\sin\theta^{\prime}\cos\theta\right]_{x=L/2},\label{eq:AppendixKL1Dim}\\
\frac{d\theta(z=+0)}{dz}=\alpha\left[\cos\theta^{\prime}\sin\theta-\cos\left(\chi-\chi^{\prime}\right)\sin\theta^{\prime}\cos\theta\right]_{x=L/2},\label{eq:AppendixKL2Dim}\\
\left.\frac{1}{\alpha^{\prime}}\sin^{2}\theta^{\prime}\frac{d\chi^{\prime}}{dy}\right|_{-0}=\left.\frac{1}{\alpha}\sin^{2}\theta\frac{d\chi}{dz}\right|_{+0}=\left.\sin\left(\chi-\chi^{\prime}\right)\sin\theta^{\prime}\sin\theta\right|_{x=L/2}.\label{eq:AppendixKL3Dim}
\end{gather}
Here the dimensionless parameters $\alpha$ and $\alpha^{\prime}$ introduced in Eq.~\eqref{eq:AlphaIneq} are assumed to be small. Physically the condition 
\begin{equation}
\alpha,\alpha^{\prime}\ll1 \, .\label{eq:AppendixAlphaSmall}
\end{equation}
follows from the condition of a weakness of the proximity effect.
Indeed, the condition~\eqref{eq:AppendixAlphaSmall} follows if the deviation
of the absolute value of the OP $|\Delta(x)|$ from the
bulk value is small which implies \eqref{eq:AppendixAlphaSmall}.
This remains correct even in the Ginzburg-Landau regime as further elaborated in Appendix~\ref{AppendixApplicability}.

We apply the perturbation theory in the small parameters, Eq.~\eqref{eq:AppendixAlphaSmall}.
This means that we are looking for the solution of the Usadel equations augmented by the boundary conditions in the form of an expansion:
\begin{gather}
\theta(z,\omega)=\theta_{0}(\omega)+\alpha\theta_{1}(z,\omega),\quad\theta^{\prime}(y,\omega)=\theta_{0}^{\prime}(\omega)+\alpha^{\prime}\theta_{1}^{\prime}(y,\omega),
\label{eq:AppendixPTTheta}\\
\Delta(z)=\Delta_{0}+\alpha\Delta_{1}(z),\quad\Delta^{\prime}(y)=\Delta_{0}^{\prime}+\alpha^{\prime}\Delta_{1}^{\prime}(y),
\label{eq:AppendixPTDelta}\\
\chi(z,\omega)=\delta\varphi+\alpha\chi_{1}(z,\omega),\quad\chi^{\prime}(y,\omega)=\alpha^{\prime}\chi_{1}^{\prime}(y,\omega),
\label{eq:AppendixPTChi}\\
\varphi(z)=\delta\varphi+\alpha\varphi_{1}(z),\quad\varphi^{\prime}(y)=\alpha^{\prime}\varphi_{1}^{\prime}(y).
\label{eq:AppendixPTVarphi}
\end{gather}
Here we fix the current-carrying state by the parameter $\delta\varphi$
phase jump of the OP $\varphi(x)$ at each of the two interfaces 
$x=\pm L/2$:
\begin{equation}
\delta\varphi\equiv\varphi(L/2)-\varphi^{\prime}(L/2)=\varphi^{\prime}(-L/2)-\varphi(-L/2).\label{eq:AppendixdeltaVarphiDef}
\end{equation}
The phase jump $\delta \varphi$ enters the system of the Usadel equations via the boundary
conditions and determines the strength of the proximity effect and
the current flowing in the system. It gives a compact expression for the current $J$ and the OP $\Delta$. Having obtained the results in terms of $\delta \varphi$, we recalculate the current-phase relation in terms of the phase change of the OP accumulated across the entire S$^{\prime}$ region. This procedure is presented in  Appendix~\ref{AppendixCPRDeltaVarphi}.

To the leading approximation, $\alpha=\alpha^{\prime}=0$ the S and S$^{\prime}$ regions are disconnected and the OPs take their respective equilibrium values $\Delta_0$ and $\Delta'_0$. The current is zero in this case. In this approximation we have for the phases
\begin{gather}
\theta(z,\omega)=\theta_{0}(\omega)\equiv\arctan\left(\frac{\Delta_{0}}{\omega}\right),\ z>0,\label{eq:AppendixSThetaZeroApp}\\
\chi(z,\omega)=\varphi(z)\equiv\delta\varphi,\label{eq:AppendixSChiZeroApp}
\end{gather}
\begin{gather}
\theta^{\prime}(y,\omega)=\theta_{0}^{\prime}(\omega),\ -L/\xi^{\prime}<y<0,\label{eq:AppendixSprThetaZeroApp}\\
\left(\Delta_{0}^{\prime}-\omega\tan\theta_{0}^{\prime}\right)^{2}=\Gamma^{\prime2}\frac{\tan^{2}\theta_{0}^{\prime}}{1+\tan^{2}\theta_{0}^{\prime}},\label{eq:AppendixSprThetaZeroApp1}\\
\chi^{\prime}(y,\omega)=\varphi^{\prime}(y)=0.\label{eq:AppendixSprChiZeroApp}
\end{gather}
With the phases $\theta$, $\theta'$ given by Eqs.~\eqref{eq:AppendixSThetaZeroApp} and \eqref{eq:AppendixSprThetaZeroApp}, and the phases $\chi$, $\chi'$ given by Eqs.~\eqref{eq:AppendixSChiZeroApp} and \eqref{eq:AppendixSprChiZeroApp} the current to the leading order is given by Eq.~\eqref{eq:AppendixCurrent}:
\begin{equation}
J_{1}(\delta\varphi)=\sin\delta\varphi\frac{2\pi T}{eR_{\text{int}}}\sum_{\omega>0}\sin\theta_{0}(\omega)\sin\theta_{0}^{\prime}(\omega).\label{eq:AppendixAmbegaokarBaratoff}
\end{equation}
In the limit of weak depairing, $\Gamma^{\prime}\ll\max\left(\Delta_{0}^{\prime},T\right)$, 
Eq.~\eqref{eq:AppendixSprThetaZeroApp1} gives 
$\tan\theta_{0}^{\prime}\approx\Delta_{0}^{\prime}/\omega$, and Eq.~\eqref{eq:AppendixAmbegaokarBaratoff} reproduces the celebrated Ambegaokar-Baratoff formula ~\eqref{eq:Current1Order}.

Notice that the calculation of the current to the first order in $\alpha$ ($\alpha'$) requires the solution of the Usadel equations to the zeroth order.
This observation is general: the calculation of the current to the $(\ell+1)$th order requires the solutions of the Usadel equations to the $\ell$th order.
Hence, below we present the solution of the Usadel equations to the first order in $\alpha$ ($\alpha'$).
This allows us to obtain the current to the second order in the same parameter(s).

\subsection{Order-parameter amplitude in the S region} \label{AppendixSRegionOrderParameter}

In this section we obtain the self-consistent expression for $\Delta_{1}(z)$
and $\theta_{1}(z,\omega)$ valid to the first order in perturbation theory. Neglecting the $\alpha^{2}$ orders in the Eqs.~\eqref{eq:Diffusion1D},
~\eqref{eq:KLR} and real part of Eq.~\eqref{eq:SelfConsistency1D},
we get the following system of equations:
\begin{gather}
\frac{d^{2}\theta_{1}(z,\omega)}{dz^{2}}+\frac{\Delta_{1}(z)}{\Delta_{0}}\cos\theta_{0}(\omega)-\frac{\theta_{1}(z)}{\sin\theta_{0}(\omega)}=0,\label{eq:AppendixSLinTheta}\\
\frac{d\theta_{1}}{dz}(z=+0)=\cos\theta_{0}^{\prime}\sin\theta_{0}-\cos\delta\varphi\sin\theta_{0}^{\prime}\cos\theta_{0},\label{eq:AppendixSLinBC}\\
\Delta_{1}(z)=\pi\lambda T\sum_{|\omega|<\omega_{D}}\theta_{1}(z,\omega)\cos\theta_{0}(\omega),\label{eq:AppendixSLinSC}
\end{gather}
The resulting system of equations is linear and translationally invariant.
To solve it, we extend $\Delta_{1}$ and $\theta_{1}$ to the region
$z<0$ to impose the boundary condition Eq.~\eqref{eq:AppendixSLinBC}
and reduce the problem to the solution of the equations with a potential
in the form of the Dirac delta function. We obtain (in what follows
we omit the dependence of $\theta_{0},\ \theta_{1}$ on $\omega$
for the sake of brevity)
\begin{equation}
\frac{d^{2}\theta_{1}}{dz^{2}}+\frac{\Delta_{1}}{\Delta_{0}}\cos\theta_{0}-\frac{\theta_{1}}{\sin\theta_{0}}=2\left(\cos\theta_{0}^{\prime}\sin\theta_{0}-\cos\delta\varphi\sin\theta_{0}^{\prime}\cos\theta_{0}\right)\delta(z).\label{eq:AppendixSLinDiracDelta}
\end{equation}
Since the system is translationally invariant, we apply the Fourier
transform to solve it [$f(k)=\intop_{-\infty}^{\infty}e^{-ikz}f(z)dz$] which gives
\begin{equation}
\theta_{1}(k)=\frac{\sin\theta_{0}}{k^{2}\sin\theta_{0}+1}\left[\frac{\Delta_{1}(k)}{\Delta_{0}}\cos\theta_{0}-2\left(\cos\theta_{0}^{\prime}\sin\theta_{0}-\cos\delta\varphi\sin\theta_{0}^{\prime}\cos\theta_{0}\right)\right].\label{eq:AppendixSLinThetaCon}
\end{equation}
We substitute the resulting relationship between $\theta_{1}$ and
$\Delta_{1}$ [Eq.~\eqref{eq:AppendixSLinThetaCon}] into the self-consistency
equation~\eqref{eq:AppendixSLinSC}. In order to eliminate the dependence
of $\theta_{1}$ and $\Delta_{1}$ on the coupling constant $\lambda$ and the high-energy cutoff $\omega_{D}$, we use the identity following from the self-consistency equation~\eqref{eq:SelfConsistency1D}
for the zero order on $\alpha$
\begin{equation}
\frac{1}{\lambda}=\frac{\pi T}{\Delta_{0}}\sum_{|\omega|<\omega_{D}}\sin\theta_{0}.\label{eq:AppendixSLambda}
\end{equation}
Using the identity~\eqref{eq:AppendixSLambda}, the self-consistency
relation~\eqref{eq:AppendixSLinSC} takes the form
\begin{equation}
\pi T\sum_{|\omega|<\omega_{D}}\left(\frac{\Delta_{1}(k)}{\Delta_{0}}\sin\theta_{0}-\theta_{1}(k)\cos\theta_{0}\right)=0.\label{eq:AppendixSCLambdaFree}
\end{equation}
Using the formula from Eq.~\eqref{eq:AppendixSLinThetaCon} in Eq.~\eqref{eq:AppendixSCLambdaFree}, the logarithmically divergent part
of the sum is reduced and the whole sum converges, and hence we can
continue the limits of summation to infinity. As a result, we obtain
\begin{equation}
\frac{\Delta_{1}(k)}{\Delta_{0}}=-2\frac{\sum_{\omega>0}\frac{\sin\theta_{0}\cos\theta_{0}}{k^{2}\sin\theta_{0}+1}\left(\cos\theta_{0}^{\prime}\sin\theta_{0}-\cos\delta\varphi\sin\theta_{0}^{\prime}\cos\theta_{0}\right)}{k^{2}\sum_{\omega>0}\frac{\sin^{2}\theta_{0}}{k^{2}\sin\theta_{0}+1}+\sum_{\omega>0}\frac{\sin^{3}\theta_{0}}{k^{2}\sin\theta_{0}+1}}.\label{eq:AppendixDelta1FourierAns}
\end{equation}
In the limit of negligible depairing we obtain the result~\eqref{eq:DeltaSimpleAns} of the main text in terms of the auxiliary functions, $L_{n,m}(k)$ introduced in Eqs.~\eqref{eq:Ldefine}.
Using the result for $\Delta_{1}(z)$ we also can find the first correction $\theta_{1}(z,\omega)$ from Eq.~\eqref{eq:AppendixSLinThetaCon} which we write here for completeness:
\begin{multline}
\theta_{1}(z>0,\omega)=-\left(\frac{\Delta_{0}}{\Delta_{0}^{\prime}}-\cos\delta\varphi\right)\left[\intop_{-\infty}^{+\infty}\frac{dk}{\pi}\cos(kz)\frac{\sin\theta_{0}(\omega)\cos\theta_{0}(\omega)}{k^{2}\sin\theta_{0}(\omega)+1}\frac{L_{1,1}(k)-L_{3,1}(k)}{k^{2}L_{2,0}(k)+L_{3,0}(k)}\right.\\
\left.-e^{-z/\sqrt{\sin\theta_{0}(\omega)}}\sqrt{\sin\theta_{0}(\omega)}\cos\theta_{0}(\omega)\sin\theta_{0}^{\prime}(\omega)\right].\label{eq:AppendixSTheta1Ans}
\end{multline}
 
\subsection{Phases of the order parameter in the S region}\label{AppendixSRegionPhases}

In this section we obtain $\varphi_{1}(z)$ and $\chi_{1}(z,\omega)$
explicitly in quadratures.~\footnote{Note that in Ref.~\cite{Osin2021}, a similar method was used to determine the phase difference, but in the end
the solution was reduced to the search for solutions of the integral
equation, which was carried out only numerically. In this paper we
give an answer how to solve such systems analytically.} 
Linearization of Eqs.~\eqref{eq:Continuity1D}, ~\eqref{eq:KLCC}, and imaginary part of~\eqref{eq:SelfConsistency1D} gives
\begin{gather}
\sin\theta_{0}(\omega)\frac{d^{2}\chi_{1}(z,\omega)}{dz^{2}}=\chi_{1}(z,\omega)-\varphi_{1}(z),\label{eq:AppendixSLinChi}\\
\frac{d\chi_{1}}{dz}(+0,\omega)=\sin\delta\varphi\frac{\sin\theta_{0}^{\prime}(\omega)}{\sin\theta_{0}(\omega)},\label{eq:AppendixSChiBC}\\
\sum_{|\omega|<\omega_{D}}\left[\chi_{1}(z,\omega)-\varphi_{1}(\omega)\right]\sin\theta_{0}(\omega)=0.\label{eq:AppendixSChiSC}
\end{gather}
In order to take advantage of the symmetrization of solutions and
Fourier transform we rewrite the obtained system of equations in terms
of new functions
\begin{equation}
\phi_{1}(z,\omega)\equiv\chi_{1}(z,\omega)-\varphi_{1}(z),\quad\Phi(z)=\varphi_{1}(z)-z\frac{d\varphi_{1}(+0)}{dz}.\label{eq:Appendixphi1Def}
\end{equation}
In terms of $\phi_{1}(z,\omega)$ and $\Phi(z) $, Eqs.~\eqref{eq:AppendixSLinChi}--\eqref{eq:AppendixSChiSC} take the form~\footnote{The convenience of the transition to new variables is due to the fact that the function $\phi_{1}(z)$ tends to zero at infinity since in
the bulk $\chi(x,\omega)=\varphi(x)$, which avoids singular parts
in the Fourier image of $\phi_{1}(k)$ (for example, Dirac delta function
$\delta(k)$). In turn, since $d\Phi/dz(z=0)=0$, we can insert the
boundary condition Eq.~\eqref{eq:AppendixSLinphi1BC} in the Usadel
equation only by breaking the derivative of the function $\phi_{1}(z)$
after symmetrization.}
\begin{gather}
\sin\theta_{0}\frac{d^{2}\phi_{1}(z)}{dz^{2}}+\sin\theta_{0}\frac{d^{2}\Phi(z)}{dz^{2}}-\phi_{1}(z)=0,\label{eq:AppendixSLinphi1}\\
\frac{d\phi_{1}}{dz}(+0)=\sin\delta\varphi\frac{\sin\theta_{0}^{\prime}}{\sin\theta_{0}}-\frac{d\varphi_{1}(+0)}{dz},\label{eq:AppendixSLinphi1BC}\\
\sum_{|\omega|<\omega_{D}}\phi_{1}(z)\sin\theta_{0}=0.\label{eq:AppendixSLinphi1SC}
\end{gather}
Extending the functions $\phi_{1}(z)$ and $\Phi(z)$ in an even way
to the region $z<0$ we get
\begin{equation}
\sin\theta_{0}\frac{d^{2}\phi_{1}(z)}{dz^{2}}+\sin\theta_{0}\frac{d^{2}\Phi(z)}{dz^{2}}-\phi_{1}(z)=2\sin\theta_{0}\frac{d\phi_{1}(+0)}{dz}\delta(z).\label{eq:AppendixSLinChiDiracDelta}
\end{equation}
Passing to the Fourier images of $\phi_{1}(z)$ and $\Phi(z)$ we
obtain the relation between these functions
\begin{equation}
\phi_{1}(k)=\frac{1}{k^{2}\sin\theta_{0}+1}\left[-k^{2}\Phi(k)\sin\theta_{0}-2\sin\theta_{0}\frac{d\phi_{1}(+0)}{dz}\right].\label{eq:AppendixSLinChiFourier}
\end{equation}
Using the expression above in the self-consistency equation~\eqref{eq:AppendixSLinphi1SC} we find $\Phi(k)$. The sum over Matsubara frequencies in Eq.~\eqref{eq:AppendixSLinphi1SC}
converges and $\omega_{D}$ can be set equal to infinity:
\begin{equation}
-k^{2}\Phi(k)=2\sum_{\omega>0}\frac{[d\phi_{1}(+0)/dz]\sin^{2}\theta_{0}}{k^{2}\sin\theta_{0}+1}/\sum_{\omega>0}\frac{\sin^{2}\theta_{0}}{k^{2}\sin\theta_{0}+1}.\label{eq:AppendixSphi1PhiCon}
\end{equation}
Dividing both parts of the equation by $ik$ the equation on $\Phi(k)$
takes the form
\begin{equation}
ik\Phi(k)=\frac{2}{ik}\sum_{\omega>0}\frac{[d\phi_{1}(+0)/dz]\sin^{2}\theta_{0}}{k^{2}\sin\theta_{0}+1}/\sum_{\omega>0}\frac{\sin^{2}\theta_{0}}{k^{2}\sin\theta_{0}+1}+C\delta(k).\label{eq:AppendixSPhiDerFourier}
\end{equation}
We determine the constant $C$ from the condition $d\Phi(+0)/dz=0$
\begin{equation}
\frac{d\Phi(+0)}{dz}=\lim_{z\to+0}\intop_{-\infty}^{+\infty}\frac{dk}{\pi}\frac{\sin kz}{k}\frac{\sum_{\omega>0}\frac{[d\phi_{1}(+0)/dz]\sin^{2}\theta_{0}}{k^{2}\sin\theta_{0}+1}}{\sum_{\omega>0}\frac{\sin^{2}\theta_{0}}{k^{2}\sin\theta_{0}+1}}+C=0.\label{eq:AppendixSZeroDer}
\end{equation}
After substituting the variable $k\to kz$ under the sign of the integral,
we obtain
\begin{equation}
\lim_{z\to+0}\intop_{-\infty}^{+\infty}\frac{dk}{\pi}\frac{\sin k}{k}\frac{\sum_{\omega>0}\frac{[d\phi_{1}(+0)/dz]\sin^{2}\theta_{0}}{(k/z)^{2}\sin\theta_{0}+1}}{\sum_{\omega>0}\frac{\sin^{2}\theta_{0}}{(k/z)^{2}\sin\theta_{0}+1}}+C=0.\label{eq:AppendixSIntegralLimit}
\end{equation}
Let us give below the asymptotic behavior of sums in the integral
in the limit $k/z\to\infty$:
\begin{equation}
\left(\frac{k}{z}\right)^{2}\sum_{\omega>0}\frac{\sin^{2}\theta_{0}}{(k/z)^{2}\sin\theta_{0}+1}\approx\sum_{\omega>0}^{1\sim(k/z)^{2}\Delta_{0}/\omega}\sin\theta_{0}\approx\frac{\Delta_{0}}{2\pi T}\left[\ln\left(\frac{k^{2}}{z^{2}}\frac{\Delta_{0}}{2\pi T}\right)+O(1)\right]\to\infty,\label{eq:AppendixSL20Limit}
\end{equation}
\begin{equation}
\lim_{k/z\to\infty}\left(\frac{k}{z}\right)^{2}\sum_{\omega>0}\frac{[d\phi_{1}(+0)/dz]\sin^{2}\theta_{0}}{(k/z)^{2}\sin\theta_{0}+1}=\lim_{k/z\to\infty}\sum_{\omega>0}\frac{d\phi_{1}(+0)}{dz}\sin\theta_{0}=0.\label{eq:AppendixSUpperSumLimit}
\end{equation}
The latter identity is a consequence of the self-consistency equation~\eqref{eq:AppendixSLinphi1SC} differentiated with respect to $z$. Thus in the limit $k/z\to\infty$ the integral is 0, and therefore $C=0$. Using
the formula~\eqref{eq:AppendixSPhiDerFourier} and the definition
of $\Phi$ [Eq.~\eqref{eq:Appendixphi1Def}] we find $\varphi_{1}(z)$:
\begin{equation}
\varphi_{1}(z>0)-\varphi_{1}(+0)=\intop_{0}^{z}dt
\intop_{-\infty}^{+\infty}\frac{dk}{\pi}\frac{\sin kz}{k}\frac{\sum_{\omega>0}\frac{[d\phi_{1}(+0)/dz]\sin^{2}\theta_{0}}{k^{2}\sin\theta_{0}+1}}{\sum_{\omega>0}\frac{\sin^{2}\theta_{0}}{k^{2}\sin\theta_{0}+1}}+z\frac{d\varphi_{1}(+0)}{dz}.\label{eq:AppendixSVarphi1Integral}
\end{equation}
Substituting the boundary condition~\eqref{eq:AppendixSLinphi1BC} into $d\phi_{1}(+0)/dz$ in the formula above the linear part $zd\varphi_{1}(+0)/dz$
is canceled and we obtain (the formula below does not imply $\Gamma^{\prime}$ to
be small) the expression in the form of Eq.~\eqref{eq:Varphi1SimpleAns} of the main text:
\begin{equation}
\varphi_{1}(z>0)-\varphi_{1}(+0)=\sin\delta\varphi\intop_{-\infty}^{\infty}\frac{dk}{\pi}\frac{1-\cos kz}{k^{2}}\frac{L_{1,1}(k)}{L_{2,0}(k)}.\label{eq:AppendixSVarphi1AnsDep}
\end{equation}
Since the phase jump of the OP is included in the definition
of $\delta\varphi$, the constant $\varphi_{1}(+0)$ is determined
from the condition of continuity of the $\varphi_{1}$ correction
on different sides of the interface, i.e., $\alpha\varphi_{1}(+0)=\alpha^{\prime}\varphi_{1}^{\prime}(-0)$.
This constant will be derived in the Appendix \ref{AppendixS'RegionPhases}.

Using the relationship between $\phi_{1}(k)$ and $\Phi(k)$ {[}Eq.~\eqref{eq:AppendixSLinChiFourier}{]} we find $\phi_{1}(z,\omega)$
{[}the function $\chi_{1}(z,\omega)$ can be found with the use of
definition of $\phi_{1}(z,\omega)$, Eq.~\eqref{eq:Appendixphi1Def}{]}:
\begin{equation}
\phi_{1}(z>0,\omega)=\sin\delta\varphi\left(\intop_{-\infty}^{+\infty}\frac{dk}{\pi}\frac{\sin\theta_{0}(\omega)\cos(kz)}{k^{2}\sin\theta_{0}(\omega)+1}\frac{L_{1,1}(k)}{L_{2,0}(k)}-\frac{\sin\theta_{0}^{\prime}(\omega)}{\sqrt{\sin\theta_{0}(\omega)}}e^{-z/\sqrt{\sin\theta_{0}(\omega)}}\right).\label{eq:AppendixSphi1AnsDep}
\end{equation}
The asymptotic behavior of the function $\varphi_{1}(z)$ {[}Eq.~\eqref{eq:AppendixSVarphi1AnsDep}{]} within $z\to0$ and $z\to\infty$ is given by the formulas
\begin{gather}
\varphi_{1}(z\to0)-\varphi_{1}(+0)\approx\sin\delta\varphi\frac{\Delta_{0}^{\prime}}{\Delta_{0}}z,\label{eq:AppendixSVarphi10Asymp}\\
\varphi_{1}(z\to\infty)\approx\sin\delta\varphi\frac{L_{1,1}(0)}{L_{2,0}(0)}z+\text{const}.\label{eq:AppendixSVarphi1InftyAsymp}
\end{gather}

The estimation of asymptotic at $z\to0$ can be done by analyzing
the ratio of sums $L_{1,1}(k)/L_{2,0}(k)$ in the limit $k\to\infty$
in a way similar to the result~\eqref{eq:AppendixSL20Limit}.
To evaluate the asymptotic at $z\to\infty$ passing to the variable
$kz\to k$, the functions $L_{n,m}(k/z)$ change slowly on the scale
of convergence of the integral and therefore the argument within the
functions can be replaced by 0. Since the derivatives of the function $\varphi_{1}(z)$
at $z=0$ and $\infty$ are different, $\varphi_{1}(z)$ behaves
nonlinearly for $z>0$.

\subsection{S$^\prime$ region: Order parameter}\label{AppendixS'RegionOrderParameter}

In this section, we find explicit expressions for $\Delta_{1}^{\prime}(y)$
and $\theta_{1}^{\prime}(y,\omega)$ inside the S$^\prime$ region, i.e., $-L/\xi^{\prime}<y<0$. To this end, 
we linearize Eqs.~\eqref{eq:Diffusion1D-1},~\eqref{eq:KLL} and the real part of Eq.~\eqref{eq:SelfConsistency1D-1} with respect to
$\alpha^{\prime}$:
\begin{gather}
\frac{d^{2}\theta_{1}^{\prime}(y,\omega)}{dy^{2}}+\frac{\Delta_{1}^{\prime}(y)}{\Delta_{0}^{\prime}}\cos\theta_{0}^{\prime}(\omega)-\frac{\theta_{1}^{\prime}(y,\omega)}{\beta(\omega)}=0,\label{eq:AppendixSPrLinTheta}\\
\frac{d\theta_{1}^{\prime}(-0,\omega)}{dy}=-\frac{d\theta_{1}^{\prime}(-L/\xi^{\prime}+0,\omega)}{dy}=\cos\delta\varphi\cos\theta_{0}^{\prime}(\omega)\sin\theta_{0}(\omega)-\sin\theta_{0}^{\prime}(\omega)\cos\theta_{0}(\omega),\label{eq:AppendixSPrLinBC}\\
\Delta_{1}^{\prime}(y)=\pi\lambda^{\prime}T\sum_{|\omega|<\omega_{D}^{\prime}}\theta_{1}^{\prime}(y,\omega)\cos\theta_{0}^{\prime}(\omega),\label{eq:AppendixSPrLinSC}
\end{gather}
where $\beta(\omega)$ is defined the following way:
\begin{equation}
\beta^{-1}(\omega)\equiv\sin\theta_{0}^{\prime}(\omega)+\frac{\omega}{\Delta_{0}^{\prime}}\cos\theta_{0}^{\prime}(\omega)+\frac{\Gamma^{\prime}}{\Delta_{0}^{\prime}}\cos2\theta_{0}^{\prime}(\omega).\label{eq:AppendixSPrfDef}
\end{equation}

The system of equations~\eqref{eq:AppendixSPrLinTheta}--\eqref{eq:AppendixSPrLinSC} is linear as in the previous cases, but is defined on the interval
$y\in\left[-L/\xi^{\prime};0\right]$. Therefore, we will use Fourier
series expansion to solve this system~\footnote{The condition $d\theta_{1}^{\prime}(-0)/dy=-d\theta_{1}^{\prime}(-L/\xi^{\prime}+0)/dy$
and the junction symmetry with respect to the point $y=-L/2\xi^{\prime}$
means that the functions $\Delta_{1}^{\prime}(y)$ and $\theta_{1}^{\prime}(y,\omega)$
are periodic with period $L/\xi^{\prime}$.}:
\begin{equation}
f(y)=\sum_{n=-\infty}^{\infty}f_{n}e^{ik_{n}y},\ k_{n}=\frac{2\pi n}{L/\xi^{\prime}},\ f_{n}=\frac{\xi^{\prime}}{L}\intop_{-L/2\xi^{\prime}}^{L/2\xi^{\prime}}dy\ f(y)e^{-ik_{n}y}.\label{eq:AppendixSPrFourierSeriesTheta}
\end{equation}

We continue the functions $\Delta_{1}^{\prime}(y)$ and $\theta_{1}^{\prime}(y,\omega)$
in a periodic manner outside the interval $y\in\left[-L/\xi^{\prime};0\right]$.
Thus, we can insert the boundary condition~\eqref{eq:AppendixSPrLinBC}
into the Usadel equation~\eqref{eq:AppendixSPrLinTheta} defined on
the entire axis
\begin{equation}
\frac{d^{2}\theta_{1}^{\prime}}{dy^{2}}+\frac{\Delta_{1}^{\prime}}{\Delta_{0}^{\prime}}\cos\theta_{0}^{\prime}-\frac{\theta_{1}^{\prime}}{\beta}=-2\frac{d\theta_{1}^{\prime}}{dy}(-0)\sum_{n=-\infty}^{\infty}\delta\left(y-\frac{L}{\xi^{\prime}}n\right).\label{eq:AppendixSPrDiracDelta}
\end{equation}

Applying the Fourier series expansion to Eq.~\eqref{eq:AppendixSPrDiracDelta}
we obtain the relationship between the Fourier coefficients
\begin{equation}
\theta_{1,n}^{\prime}=\frac{\beta}{k_{n}^{2}\beta+1}\left[\frac{\Delta_{1,n}^{\prime}}{\Delta_{0}^{\prime}}\cos\theta_{0}^{\prime}+2\frac{\xi^{\prime}}{L}\frac{d\theta_{1}^{\prime}(-0)}{dy}\right].\label{eq:AppendixSPrTheta1Con}
\end{equation}
Similarly as we obtained the relation~\eqref{eq:AppendixDelta1FourierAns}
we find the Fourier coefficients for $\Delta_{1}^{\prime}(y)$ with
the use of the self-consistency equation~\eqref{eq:AppendixSPrLinSC}
and the connection between $\theta_{1,n}^{\prime}$ and $\Delta_{1,n}^{\prime}$:
\begin{gather}
\frac{1}{\lambda^{\prime}}=\frac{\pi T}{\Delta_{0}^{\prime}}\sum_{|\omega|<\omega_{D}^{\prime}}\sin\theta_{0}^{\prime},\label{eq:AppendixSPrLambda}\\
\sum_{|\omega|<\omega_{D}^{\prime}}\left(\frac{\Delta_{1,n}^{\prime}}{\Delta_{0}^{\prime}}\sin\theta_{0}^{\prime}-\theta_{1,n}^{\prime}\cos\theta_{0}^{\prime}\right)=0\label{eq:AppendixSPrSCLambdaFree}\\
\Rightarrow\frac{\Delta_{1,n}^{\prime}}{\Delta_{0}^{\prime}}=2\frac{\xi^{\prime}}{L}\sum_{\omega>0}\frac{\beta\cos\theta_{0}^{\prime}d\theta_{1}^{\prime}(-0)/dy}{k_{n}^{2}\beta+1}
\left(k_{n}^{2}\sum_{\omega>0}\frac{\beta\sin\theta_{0}^{\prime}}{k_{n}^{2}\beta+1}+\sum_{\omega>0}\frac{\sin\theta_{0}^{\prime}-\beta\cos^{2}\theta_{0}^{\prime}}{k_{n}^{2}\beta+1}\right)^{-1}.\label{eq:AppendixSPrDelta1Fourier}
\end{gather}
In the limit $\Gamma^{\prime}\ll\max\left(\Delta_{0}^{\prime},T\right)$,
$\beta\approx\sin\theta_{0}^{\prime}$ and we obtain for $\Delta_{1}^{\prime}(y)$
and $\theta_{1}^{\prime}(y,\omega)$ {[}see the definition of $L_{n,m}^{\prime}$
sums, Eq.~\eqref{eq:Ldefine}{]}
\begin{gather}
\frac{\Delta_{1}^{\prime}(y)}{\Delta_{0}^{\prime}}=-2\frac{\xi^{\prime}}{L}\left(\frac{\Delta_{0}^{\prime}}{\Delta_{0}}-\cos\delta\varphi\right)\sum_{n=-\infty}^{\infty}\cos(k_{n}y)\frac{L_{1,1}^{\prime}(k_{n})-L_{3,1}^{\prime}(k_{n})}{k_{n}^{2}L_{2,0}^{\prime}(k_{n})+L_{3,0}^{\prime}(k_{n})},\ y\in\left[-L/\xi^{\prime};0\right],\label{eq:AppendixSPrDelta1Ans}
\end{gather}
\begin{multline}
\theta_{1}^{\prime}(y,\omega)=\theta_{1}^{\prime}\left(-L/\xi^{\prime}-y,\omega\right)=\sum_{n=-\infty}^{\infty}\frac{\sin\theta_{0}^{\prime}(\omega)\cos\theta_{0}^{\prime}(\omega)}{k_{n}^{2}\sin\theta_{0}^{\prime}(\omega)+1}\frac{\Delta_{1,n}^{\prime}}{\Delta_{0}^{\prime}}\cos(k_{n}y)\\
-\left(\frac{\Delta_{0}^{\prime}}{\Delta_{0}}-\cos\delta\varphi\right)\sqrt{\sin\theta_{0}^{\prime}(\omega)}\cos\theta_{0}^{\prime}(\omega)\sin\theta_{0}(\omega)\frac{\cosh\left[(L/2\xi^{\prime}+y)/\sqrt{\sin\theta_{0}^{\prime}(\omega)}\right]}{\sinh\left[L/2\xi^{\prime}\sqrt{\sin\theta_{0}^{\prime}(\omega)}\right]},\\
y\in\left[-L/2\xi^{\prime};0\right].\label{eq:AppendixSPrTheta1Ans}
\end{multline}
When deriving $\theta_{1}^{\prime}(y,\omega)$ we have used the formula
\begin{equation}
\sum_{n=-\infty}^{\infty}\frac{e^{in\gamma}}{n^{2}+a^{2}}=\frac{\pi\cosh\left[a\left(\pi-|\gamma|\right)\right]}{a\sinh\pi a},\ \gamma\in\left[-\pi;\pi\right],\ a\in\mathbb{R}.\label{eq:AppendixSPrSum1}
\end{equation}

\subsection{S$^\prime$ region: Phases}\label{AppendixS'RegionPhases}

In this section we derive analytic expressions for $\varphi_{1}^{\prime}(y)$
and $\chi_{1}^{\prime}(y,\omega)$. We linearize Eqs.~\eqref{eq:Continuity1D-1},~\eqref{eq:KLCC}, and the imaginary part of Eq.~\eqref{eq:SelfConsistency1D-1}
with respect to $\alpha^{\prime}$:
\begin{gather}
\sin\theta_{0}^{\prime}\frac{d^{2}\chi_{1}^{\prime}(y,\omega)}{dy^{2}}=\chi_{1}^{\prime}(y,\omega)-\varphi_{1}^{\prime}(y),\label{eq:AppendixSPrLinChi}\\
\frac{d\chi_{1}^{\prime}(-0,\omega)}{dy}=\frac{d\chi_{1}^{\prime}(-L/\xi^{\prime}+0,\omega)}{dy}=\sin\delta\varphi\frac{\sin\theta_{0}(\omega)}{\sin\theta_{0}^{\prime}(\omega)},\label{eq:AppendixSPrLinChiBC}\\
\sum_{|\omega|<\omega_{D}^{\prime}}\left[\chi_{1}^{\prime}(y,\omega)-\varphi_{1}^{\prime}(y)\right]\sin\theta_{0}^{\prime}(\omega)=0.\label{eq:AppendixSPrLinChiSC}
\end{gather}

Due to the symmetry~\footnote{Equations~\eqref{eq:AppendixSPrLinChi}--\eqref{eq:AppendixSPrLinChiSC}
allow searching for solutions in the form of odd functions $f(-L/2\xi^{\prime}-y)=-f(-L/2\xi^{\prime}+y)$.} of the system with respect to $y=-L/2\xi^{\prime}$ we set $\chi_{1}^{\prime}(-L/2\xi^{\prime},\omega)=\varphi_{1}^{\prime}(-L/2\xi^{\prime})=0$.
We define functions $\Phi^{\prime}(y)$ and $\phi_{1}^{\prime}(y,\omega)$
as
\begin{gather}
\phi_{1}^{\prime}(y,\omega)\equiv\chi_{1}^{\prime}(y,\omega)-\varphi_{1}^{\prime}(y),\ \phi_{1}^{\prime}(-L/2\xi^{\prime},\omega)=0,\ y\in\left[-L/\xi^{\prime};0\right],\label{eq:AppendixSPrphi1Def}\\
\Phi^{\prime}(y)=\Phi^{\prime}(-L/\xi^{\prime}-y)=\varphi_{1}^{\prime}(y)-y\frac{d\varphi_{1}^{\prime}(-0)}{dy}.\label{eq:AppendixSPrVarphiDef}
\end{gather}
In terms of new variables, Eqs.~\eqref{eq:AppendixSPrLinChi}--\eqref{eq:AppendixSPrLinChiSC}
take the form
\begin{gather}
\sin\theta_{0}^{\prime}\frac{d^{2}\phi_{1}^{\prime}(y,\omega)}{dy^{2}}+\sin\theta_{0}^{\prime}\frac{d^{2}\Phi^{\prime}(y)}{dy^{2}}=\phi_{1}^{\prime}(y,\omega),\label{eq:AppendixSPrLinphi1}\\
\frac{d\phi_{1}^{\prime}(-0,\omega)}{dy}=\frac{d\phi_{1}^{\prime}(-L/\xi^{\prime}+0,\omega)}{dy}=\sin\delta\varphi\frac{\sin\theta_{0}(\omega)}{\sin\theta_{0}^{\prime}(\omega)}-\frac{d\varphi_{1}^{\prime}(-0)}{dy},\label{eq:AppendixSPrLinphi1BC}\\
\sum_{|\omega|<\omega_{D}^{\prime}}\phi_{1}^{\prime}(y,\omega)\sin\theta_{0}^{\prime}(\omega)=0.\label{eq:AppendixSPrLinphi1SC}
\end{gather}
Due to relation $d\chi_{1}^{\prime}(-0,\omega)/dy=d\chi_{1}^{\prime}(-L/\xi^{\prime}+0,\omega)$
and the symmetry of the system with respect to $y=-L/2\xi^{\prime}$,
the period of functions $\phi_{1}^{\prime}(y,\omega)$ and $\Phi^{\prime}(y)$
extended in the even way with respect to $y=0$ over the interval
$[-L/\xi^{\prime};0]$ is equal to $2L/\xi^{\prime}$ . Therefore,
the Fourier series expansion will be carried out in wave vectors $q_{n}$
half as large as $k_{n}$:
\begin{equation}
f(y)=\sum_{n=-\infty}^{\infty}f_{n}e^{iq_{n}y},\ q_{n}=\frac{\pi n}{L/\xi^{\prime}},\ f_{n}=\frac{\xi^{\prime}}{2L}\intop_{-L/\xi^{\prime}}^{L/\xi^{\prime}}dyf(y)e^{-iq_{n}y}.\label{eq:AppendixSPrFourierSeriesChi}
\end{equation}
We insert the boundary condition~\eqref{eq:AppendixSPrLinphi1BC}
into the Usadel equation~\eqref{eq:AppendixSPrLinphi1} using the
Dirac delta function
\begin{equation}
\sin\theta_{0}^{\prime}\frac{d^{2}\phi_{1}^{\prime}}{dy^{2}}+\sin\theta_{0}^{\prime}\frac{d^{2}\Phi^{\prime}}{dy^{2}}=\phi_{1}^{\prime}-2\frac{d\phi_{1}^{\prime}(-0)}{dy}\sin\theta_{0}^{\prime}\sum_{n=-\infty}^{\infty}(-1)^{n}\delta\left(y-\frac{2L}{\xi^{\prime}}n\right).\label{eq:AppendixSPrphi1DiracDelta}
\end{equation}
Applying the Fourier series expansion to the resulting equation, we
obtain a connection between the Fourier coefficients $\phi_{1,n}^{\prime}$
and $\Phi_{n}^{\prime}$:
\begin{equation}
\phi_{1,n}^{\prime}=\frac{\sin\theta_{0}^{\prime}}{q_{n}^{2}\sin\theta_{0}^{\prime}+1}\left\{ -q_{n}^{2}\Phi_{n}^{\prime}+\frac{d\phi_{1}^{\prime}(-0)}{dy}\cdot\frac{\xi^{\prime}}{L}\left[1-(-1)^{n}\right]\right\} .\label{eq:AppendixSPrphi1Con}
\end{equation}
Substituting the above relation into the self-consistency equation~\eqref{eq:AppendixSPrLinphi1SC} we find the coefficients $\Phi_{n}^{\prime}$
(the formula below does not imply $\Gamma^{\prime}$ to be small):
\begin{equation}
\Phi_{n}^{\prime}=\frac{\xi^{\prime}}{q_{n}^{2}L}\left[1-(-1)^{n}\right]\left\{ \sin\delta\varphi\frac{L_{1,1}^{\prime}(q_{n})}{L_{2,0}^{\prime}(q_{n})}-\frac{d\varphi_{1}^{\prime}(-0)}{dy}\right\} ,\ \Phi_{0}^{\prime}=0.\label{eq:AppendixSPrPhiFourierCoeff}
\end{equation}
Thus, we find the expression for $\varphi_{1}^{\prime}(y)$ presented in the main text as Eq.~\eqref{eq:VarphiPrSimpleAns}. 
Since in Eq.~\eqref{eq:AppendixSPrPhiFourierCoeff} $n$ only runs
through odd values, it is more convenient to proceed to summation
over wave vectors $k_{n+1/2}$ {[}Eq.~\eqref{eq:AppendixSPrFourierSeriesTheta}{]}:
\begin{gather}
\varphi_{1}^{\prime}(y)=\sin\delta\varphi\frac{2\xi^{\prime}}{L}\sum_{n=-\infty}^{\infty}\frac{\cos(k_{n+1/2}y)-1}{k_{n+1/2}^{2}}\frac{L_{1,1}^{\prime}(k_{n+1/2})}{L_{2,0}^{\prime}(k_{n+1/2})}+\varphi_{1}^{\prime}(-0),\label{eq:AppendixSPrvarphi1Ans}\\
\varphi_{1}^{\prime}(-0)=\sin\delta\varphi\frac{2\xi^{\prime}}{L}\sum_{n=-\infty}^{\infty}\frac{1}{k_{n+1/2}^{2}}\frac{L_{1,1}^{\prime}(k_{n+1/2})}{L_{2,0}^{\prime}(k_{n+1/2})}.\label{eq:AppendixSPrvarphi1Interface}
\end{gather}
The last expression allows us to calculate the constant $\varphi_{1}(z=+0) ${[}Eq.~\eqref{eq:AppendixSVarphi1AnsDep}{]}:
\begin{equation}
\varphi_{1}(z=+0)=\frac{\alpha^{\prime}}{\alpha}\varphi_{1}^{\prime}(y=-0).\label{eq:AppendixSPrVarphi1(+0)}
\end{equation}
We obtain the function $\phi_{1}^{\prime}(y,\omega)$ from Eq.~\eqref{eq:AppendixSPrPhiFourierCoeff} {[}the function $\chi_{1}^{\prime}(y,\omega)$
can be found with the use of definition of $\phi_{1}^{\prime}(y,\omega)$,
Eq.~\eqref{eq:AppendixSPrphi1Def}{]}:
\begin{multline}
\phi_{1}^{\prime}(y,\omega)=-\sin\delta\varphi\frac{2\xi^{\prime}}{L}\sum_{n=-\infty}^{\infty}\frac{\sin\theta_{0}^{\prime}(\omega)}{k_{n+1/2}^{2}\sin\theta_{0}^{\prime}(\omega)+1}\frac{L_{1,1}^{\prime}(k_{n+1/2})}{L_{2,0}^{\prime}(k_{n+1/2})}\cos(k_{n+1/2}y)\\
+\sin\delta\varphi\frac{\sin\theta_{0}(\omega)}{\sqrt{\sin\theta_{0}^{\prime}(\omega)}}\frac{\sinh\left[\left(y+L/2\xi^{\prime}\right)/\sqrt{\sin\theta_{0}^{\prime}(\omega)}\right]}{\cosh\left(L/2\xi^{\prime}\sqrt{\sin\theta_{0}^{\prime}(\omega)}\right)}.\label{eq:AppendixSPrphi1Ans}
\end{multline}
When deriving expressions in Eqs.~\eqref{eq:AppendixSPrvarphi1Ans} and~\eqref{eq:AppendixSPrphi1Ans}
we used the relation
\begin{equation}
\sum_{n=-\infty}^{\infty}\frac{e^{i(2n+1)\gamma}}{(2n+1)^{2}+a^{2}}=\frac{\pi}{2a}\frac{\sinh\left[\left(\frac{\pi}{2}-|\gamma|\right)a\right]}{\cosh\left(\pi a/2\right)},\ \gamma\in\left[-\pi;\pi\right].\label{eq:AppendixSPrSum2}
\end{equation}


\section{Josephson current}\label{AppendixCPR}

In this appendix we derive the current-phase relation taking
into account first-order perturbation theory corrections with respect
to $\alpha$ and $\alpha^{\prime}$ for $\Delta$ and $\varphi$ in
the S and S$^\prime$ regions for the case of small $\Gamma^{\prime}\ll\max\left(\Delta_{0}^{\prime},T\right)$ [see
Eqs.~\eqref{eq:DeltaSimpleAns},~\eqref{eq:Varphi1SimpleAns},~\eqref{eq:AppendixSPrDelta1Ans}, and~\eqref{eq:AppendixSPrvarphi1Ans}].

We use Eq.~\eqref{eq:AppendixCurrent} and expand it to terms
of order $\alpha^{2},\alpha^{\prime2}$:
\begin{multline}
J(\delta\varphi)=\frac{\pi T}{eR_{\text{int}}}\sum_{\omega}\sin\left(\delta\varphi+\alpha\phi_{1}(0)-\alpha^{\prime}\phi_{1}^{\prime}(0)\right)\sin\left[\theta_{0}+\alpha\theta_{1}(0)\right]\sin\left[\theta_{0}^{\prime}+\alpha^{\prime}\theta_{1}^{\prime}(0)\right]\\
=J_{1}(\delta\varphi)+\frac{\pi T}{eR_{\text{int}}}\cos\delta\varphi\sum_{\omega}\sin\theta_{0}\sin\theta_{0}^{\prime}\left[\alpha\phi_{1}(0)-\alpha^{\prime}\phi_{1}^{\prime}(0)\right]\\
+\frac{\pi T}{eR_{\text{int}}}\alpha\sin\delta\varphi\sum_{\omega}\cos\theta_{0}\sin\theta_{0}^{\prime}\theta_{1}(0)+\frac{\pi T}{eR_{\text{int}}}\alpha^{\prime}\sin\delta\varphi\sum_{\omega}\sin\theta_{0}\cos\theta_{0}^{\prime}\theta_{1}^{\prime}(0).\label{eq:AppendixCurrentExpansion}
\end{multline}

Applying Eqs.~\eqref{eq:AppendixSTheta1Ans},~\eqref{eq:AppendixSphi1AnsDep},~\eqref{eq:AppendixSPrTheta1Ans}, and~\eqref{eq:AppendixSPrphi1Ans} we
obtain the result for the current in the form~\eqref{eq:resJ}.
The leading contribution $J_{1}(\delta\varphi)$ is given by Eq.~\eqref{eq:Current1Order} and $J_{2}^{(1,2)}$ are presented in Eq.~\eqref{eq:J2Ans} of the main text, respectively.

\subsection{The sign of the correction to the first harmonic}
\label{app:signJ12}

The sign of the correction, Eq.~\eqref{eq:J2Ans}, is opposite to the main contribution for any choice of parameters such as temperature, the ratio $\Delta_{0}/\Delta_{0}^{\prime}$, and length of the junction.
Notice that according to the definitions \eqref{eq:Ldefine},
\begin{gather}
\frac{\left[L_{1,1}(k)-L_{3,1}(k)\right]^{2}}{k^{2}L_{2,0}(k)+L_{3,0}(k)}>0,\label{eq:AppendixIneq1}\\
L_{1,2}(k)-L_{3,2}(k)=\frac{2\pi T}{\Delta_{0}}\sum_{\omega>0}\frac{\sin\theta_{0}\cos^{2}\theta_{0}\sin^{2}\theta_{0}^{\prime}}{k^{2}\sin\theta_{0}+1}>0,\label{eq:AppendixIneq2}
\end{gather}
with the same conclusion for $L_{n,m}^{\prime}$ functions.
From these inequalities it follows that $S(k),S^{\prime}(k)>0$.
Therefore, the amplitude of the first harmonic correction $J_{2}^{(1)}(\delta\varphi)$
is always negative.

\subsection{Current-phase relation in terms of phase jump over the S$^\prime$ region}\label{AppendixCPRDeltaVarphi}

Using Eq.~\eqref{eq:AppendixSPrvarphi1Ans} for $\varphi_{1}^{\prime}(y)$
we find $\varphi_{\text{cur}}$:
\begin{multline}
\varphi_{\text{cur}}=\alpha^{\prime}\left[\varphi_{1}^{\prime}(0)-\varphi_{1}^{\prime}(-L/\xi^{\prime})\right]=\alpha^{\prime}\sin\delta\varphi\frac{2\xi^{\prime}}{L} \sum_{n=-\infty}^{\infty}\frac{1-\cos\left(k_{n+1/2}L/\xi^{\prime}\right)}{k_{n+1/2}^{2}}\frac{L_{1,1}^{\prime}(k_{n+1/2})}{L_{2,0}^{\prime}(k_{n+1/2})} \equiv
\alpha_{L}^{\prime}\sin\delta\varphi R(L/\xi^{\prime}),\label{eq:AppendixVarphiCurAns}
\end{multline}
with, $\alpha_{L}^{\prime}=g_{\text{int}}L/\sigma^{\prime}$, where we defined dimensionless function 
\begin{equation}
R(t)\equiv\frac{8}{\pi^{2}}\sum_{n=0}^{\infty}\frac{1}{(2n+1)^{2}}\frac{L_{1,1}^{\prime}[\pi(2n+1)/t]}{L_{2,0}^{\prime}[\pi(2n+1)/t]}.\label{eq:AppendixRDef}
\end{equation}

In order to rewrite the current-phase relation in terms
of $\Delta\varphi$ we find the solution of Eq.~\eqref{eq:AppendixDeltaVarphiDef}, $\delta\varphi(\Delta\varphi)$. We consider the tunneling limit, which
assumes that the interface resistance is the largest scale resistance
in the system {[}see Eq.~\eqref{eq:AppendixAlphaSmall}{]}, in particular
compared to the S$^\prime$ region resistance, which means $\alpha_{L}^{\prime}\ll1$.
Up to first order in $\alpha_{L}^{\prime}$ included the solution
of Eq.~\eqref{eq:AppendixDeltaVarphiDef} have the form

\begin{equation}
\delta\varphi=\frac{\Delta\varphi}{2}-\frac{R(L/\xi^{\prime})}{2}\alpha_{L}^{\prime}(-1)^{n}\sin\left(\frac{\Delta\varphi}{2}\right)-\pi n.\label{eq:AppendixdeltaVarphiConn}
\end{equation}

Since we are looking for the current-phase relation up to $\alpha^{2}$ (or
equivalently $1/R_{\text{int}}^{2}$) the correction term with $\alpha_{L}^{\prime}$
in Eq.~\eqref{eq:AppendixdeltaVarphiConn} contribute to the correction to
the first harmonic and second harmonic only in $\alpha^{3}$ ($1/R_{\text{int}}^{3}$).
Therefore, we should encounter this correction only in the main order
of the current-phase relation {[}Ambegaokar-Baratoff formula~\eqref{eq:AppendixAmbegaokarBaratoff}{]}
\begin{equation}
J_{1}(\delta\varphi)=\sin\delta\varphi\frac{2\pi T}{eR_{\text{int}}}\sum_{\omega>0}\sin\theta_{0}^{\prime}\sin\theta_{0}=
\left[(-1)^{n}\sin\left(\frac{\Delta\varphi}{2}\right)-\frac{R(L/\xi^{\prime})}{4}\alpha_{L}^{\prime}\sin\left(\Delta\varphi\right)\right]\frac{\Delta_{0}}{eR_{\text{int}}}L_{1,1}(0).\label{eq:AppendixABCorr}
\end{equation}

Thus, we obtain the current phase relationship $J(\Delta\varphi)$ in the form of Eq.~\eqref{eq:AppendixCPRDeltaVarphiAns}
with the separate contributions the functions $J_{1}(\Delta\varphi)$, $J_{2}^{(1)}(\Delta\varphi)$
and $J_{2}^{(2)}(\Delta\varphi)$ given by Eq.~\eqref{eq:JDelta12}.

\section{Asymptotic behavior}\label{AppendixAsymptotes}

In this appendix we present the asymptotic behavior for the functions
$\Delta$, $\varphi$, $J_{2}^{(1)}(\Delta\varphi)$, and $J_{2}^{(2)}(\Delta\varphi)$
as a function of junction length $L$ and temperature $T$.

\subsection{Long-junction approximation}\label{AppendixLongJunction}

In this section we consider the case $L\gg\xi^{\prime}$, $\Delta_{0},\Delta_{0}^{\prime}\gtrsim T$.
Consider the functions $\Delta_{1}^{\prime}(y)$ and $\varphi_{1}^{\prime}(y)$ {[}see Eqs.~\eqref{eq:AppendixSPrDelta1Ans} and~\eqref{eq:AppendixSPrvarphi1Ans}{]}.
In this approximation, one can move from summation over wave vectors
$k_{n}$ to integration over a continuous variable $k$. As a result,
the equations for $\Delta_{1}^{\prime}(y)$ and $\varphi_{1}^{\prime}(y)$
take the form~\eqref{eq:DeltaPrSimpleAns} and~\eqref{eq:AppendixSVarphi1AnsDep}, respectively, up to replacing $\Delta_{0}$ with $\Delta_{0}^{\prime}$
and vice versa:
\begin{gather}
\frac{\Delta_{1}^{\prime}(y<0)}{\Delta_{0}^{\prime}}=-\left(\frac{\Delta_{0}^{\prime}}{\Delta_{0}}-\cos\delta\varphi\right)\intop_{-\infty}^{+\infty}\frac{dk}{\pi}\cos(ky)\frac{L_{1,1}^{\prime}(k)-L_{3,1}^{\prime}(k)}{k^{2}L_{2,0}^{\prime}(k)+L_{3,0}^{\prime}(k)},\label{eq:AppendixLongDeltaPr}\\
\varphi_{1}^{\prime}(y<0)-\varphi_{1}^{\prime}(0)=\sin\delta\varphi\intop_{-\infty}^{+\infty}\frac{dk}{\pi}\frac{\cos ky-1}{k^{2}}\frac{L_{1,1}^{\prime}(k)}{L_{2,0}^{\prime}(k)}.\label{eq:AppendixLongVarphiPr}
\end{gather}
In a similar way we get the answer for the current-phase relation
corrections~\eqref{eq:AppendixJ21DeltaVarphiAns} and~\eqref{eq:AppendixJ2DeltaVarphiAns}:
\begin{gather}
J_{2}^{(1)}(\Delta\varphi)=-\frac{\sin\left(\Delta\varphi/2\right)\sgn\left[\cos\left(\Delta\varphi/2\right)\right]}{eR_{\text{int}}}\left[\alpha\frac{\Delta_{0}^{2}}{\Delta_{0}^{\prime}}\intop_{\mathbb{-\infty}}^{+\infty}\frac{dk}{\pi}S(k)+\alpha^{\prime}\frac{\Delta_{0}^{\prime2}}{\Delta_{0}}\intop_{\mathbb{-\infty}}^{+\infty}\frac{dk}{\pi}S^{\prime}(k)\right],\label{eq:AppendixLongJ21}
\end{gather}
\begin{equation}
J_{2}^{(2)}(\Delta\varphi)=\frac{\sin\Delta\varphi}{2eR_{\text{int}}}\left\{ \alpha\Delta_{0}\intop_{\mathbb{-\infty}}^{+\infty}\frac{dk}{\pi}\left[S(k)+P(k)\right]+\alpha^{\prime}\Delta_{0}^{\prime}\intop_{\mathbb{-\infty}}^{+\infty}\frac{dk}{\pi}\left[S^{\prime}(k)+P^{\prime}(k)\right]-\frac{\alpha_{L}^{\prime}}{2}\cdot\Delta_{0}L_{1,1}(0)\frac{L_{1,1}^{\prime}(0)}{L_{2,0}^{\prime}(0)}\right\} .\label{eq:AppendixLongJ22}
\end{equation}
Here we have used that the function $R(t)$ has a finite limit for
$t\to\infty$:
\begin{gather}
R(t\to\infty)=\frac{8}{\pi^{2}}\sum_{n=0}^{\infty}\frac{1}{(2n+1)^{2}}\frac{L_{1,1}^{\prime}(0)}{L_{2,0}^{\prime}(0)}=\frac{L_{1,1}^{\prime}(0)}{L_{2,0}^{\prime}(0)}.\label{eq:AppendixR(t)Limit}
\end{gather}

\subsection{Short-junction approximation}\label{AppendixShortJunction}

Consider the limit $L/\xi^{\prime}\ll1$, $\Delta_{0},\Delta_{0}^{\prime}\gtrsim T$.
When considering Eq.~\eqref{eq:AppendixSPrDelta1Ans} on $\Delta_{1}^{\prime}(y)$
in the sum over wave vectors, all $k_{n}\gg1$ except $k_{0}=0$.
In this case, the asymptotics of the answer is given only by the sum
term with $k=0$:
\begin{equation}
\frac{\Delta_{1}^{\prime}(y)}{\Delta_{0}^{\prime}}=-\left(\frac{\Delta_{0}^{\prime}}{\Delta_{0}}-\cos\delta\varphi\right)\frac{2\xi^{\prime}}{L}\frac{L_{1,1}^{\prime}(0)-L_{3,1}^{\prime}(0)}{L_{3,0}^{\prime}(0)}.\label{eq:AppendixShortDeltaAns}
\end{equation}
When calculating the function $\varphi_{1}^{\prime}(y)$, Eq.~\eqref{eq:AppendixSPrvarphi1Ans},
since the wave vector $k=0$ is absent in the sum, we set the argument
inside the functions $L_{n,m}^{\prime}(k)$ equal to infinity to calculate
the asymptotics
\begin{equation}
\varphi_{1}^{\prime}(y)\approx\sin\delta\varphi\frac{2\xi^{\prime}}{L}\sum_{n=-\infty}^{\infty}\frac{\cos\left(k_{n+1/2}y\right)}{k_{n+1/2}^{2}}\frac{L_{1,1}^{\prime}(\infty)}{L_{2,0}^{\prime}(\infty)}=\sin\delta\varphi\left(y+\frac{L}{2\xi^{\prime}}\right)\frac{L_{1,1}^{\prime}(\infty)}{L_{2,0}^{\prime}(\infty)}.\label{eq:AppendixShortVarphi1Approx}
\end{equation}
In the last formula we have used Eq.~\eqref{eq:AppendixSPrSum2}.
Performing operations similar to the derivation of Eq.~\eqref{eq:AppendixSL20Limit},
we obtain 
\begin{equation}
\varphi_{1}^{\prime}(y)=\sin\delta\varphi\frac{\Delta_{0}}{\Delta_{0}^{\prime}}\left(y+\frac{L}{2\xi^{\prime}}\right).\label{eq:AppendixShortVarphi1Ans}
\end{equation}
In the current-phase relation corrections $J_{2}^{(1)}(\Delta\varphi)$
and $J_{2}^{(2)}(\Delta\varphi)$, the main contribution to the leading order in terms of $\xi^{\prime}/L\gg1$ comes from the term $S^{\prime}(0)$
in the summation over wave vectors, which gives the answer in the form
\begin{gather}
J_{2}^{(1)}(\Delta\varphi)=-\frac{\sin\left(\Delta\varphi/2\right)\sgn\left[\cos\left(\Delta\varphi/2\right)\right]}{eR_{\text{int}}}\alpha^{\prime}\frac{\Delta_{0}^{\prime2}}{\Delta_{0}}\frac{2\xi^{\prime}}{L}S^{\prime}(0),\label{eq:AppendixShortJ21Ans}\\
J_{2}^{(2)}(\Delta\varphi)=\frac{\sin\Delta\varphi}{2eR_{\text{int}}}\alpha^{\prime}\Delta_{0}^{\prime}\frac{2\xi^{\prime}}{L}S^{\prime}(0).\label{eq:AppendixShortJ22Ans}
\end{gather}
\subsection{Ginzburg-Landau regime}\label{AppendixGL}

In this section we consider the limit when one the S, S$^\prime$ regions or
both at same time are near their critical temperature $T\to T_{c}(T_{c}^{\prime})$. We consider each case separately, taking into account the dependence on the junction length $L$.

\subsubsection{\texorpdfstring{$T\to T_{c}<T_{c}^{\prime},\ L\gg\xi^{\prime}$}{}}

In this limit $\Delta_{0}\ll T$ and we can use the standard equation
for $\Delta_{0}$ from BCS theory (see Ref.~\cite{Svidzinsky1982}) (the same equations can be written for $\Delta_{0}^{\prime}$ and
$\sin\theta_{0}^{\prime}$ in the limit $T\to T_{c}^{\prime}$)

\begin{equation}
\Delta_{0}^{2}(T)=\frac{8\pi^{2}T_{c}^{2}}{7\zeta(3)}\left(1-\frac{T}{T_{c}}\right),\ \sin\theta_{0}\approx\frac{\Delta_{0}}{|\omega|}.\label{eq:AppendixDelta0GL}
\end{equation}

We start from finding the functions $\Delta_{1}(z)$. Since in formula~\eqref{eq:DeltaSimpleAns} the sums $L_{n,m}(k)$ vary on a
scale of order $k\sim\left(\omega/\Delta_{0}\right)^{1/2}\gg1$, while
the integral converges on a much smaller scale $k\sim\left(L_{30}(0)/L_{20}(0)\right)^{1/2}\sim\left(\Delta_{0}/\omega\right)^{1/2}\ll1$,
we can put the argument $k$ of the functions $L_{n,m}(k)$ equal
to 0. We obtain
\begin{equation}
\frac{\Delta_{1}(z>0)}{\Delta_{0}}\approx-\left(\frac{\Delta_{0}}{\Delta_{0}^{\prime}}-\cos\delta\varphi\right)\frac{L_{11}(0)}{\sqrt{L_{20}(0)L_{30}(0)}}\exp\left(-z\sqrt{\frac{L_{30}(0)}{L_{20}(0)}}\right).\label{eq:AppendixSGLDelta1Ans1}
\end{equation}
Here we give an approximate expressions for the sums $L_{n,m}$ in
the formula above (the same kind of evaluation is true for $L_{n,m}^{\prime}$
sums in the limit $T\to T_{c}^{\prime}$):
\begin{gather}
L_{a,0}(0)=\frac{2\pi T}{\Delta_{0}}\sum_{\omega>0}\sin^{a}\theta_{0}\approx\frac{2\pi T}{\Delta_{0}}\sum_{\omega>0}\frac{\Delta_{0}^{a}}{\omega^{a}}=2\left(\frac{\Delta_{0}}{\pi T}\right)^{a-1}\left(1-\frac{1}{2^{a}}\right)\zeta(a),\label{eq:AppendixLn0Approx}\\
L_{a,b}(0)=\frac{2\pi T}{\Delta_{0}}\sum_{\omega>0}\sin^{a}\theta_{0}\sin^{b}\theta_{0}^{\prime}=2\left(\frac{\Delta_{0}}{\pi T}\right)^{a-1}\sum_{n=0}^{\infty}\left(\frac{1}{2n+1}\right)^{a}\sin^{b}\theta_{0}^{\prime}(\omega_{n}).\label{eq:AppendixLnmApprox}
\end{gather}
For convenience, we introduce dimensionless parameters and variables
that do not depend on temperature $T$:
\begin{gather}
\xi_{1}\equiv\sqrt{\frac{D}{2\pi T_{c}}},\ \xi_{1}^{\prime}\equiv\sqrt{\frac{D^{\prime}}{2\pi T_{c}}},\ \xi_{2}\equiv\sqrt{\frac{D}{2\pi T_{c}^{\prime}}},\ \xi_{2}^{\prime}\equiv\sqrt{\frac{D^{\prime}}{2\pi T_{c}^{\prime}}},\label{eq:AppendixXi12Def}\\
\alpha_{1,2}\equiv\frac{g_{\text{int}}\xi_{1,2}}{\sigma},\ \alpha_{1,2}^{\prime}\equiv\frac{g_{\text{int}}\xi_{1,2}^{\prime}}{\sigma^{\prime}},\label{eq:AppendixAlpha12Def}\\
z_{1,2}\equiv\frac{x-L/2}{\xi_{1,2}},\ y_{1,2}\equiv\frac{x-L/2}{\xi_{1,2}^{\prime}}.\label{eq:AppendixZY12Def}
\end{gather}
Thus, neglecting higher orders of $\Delta_{0}\ll T_{c}$ the
answer for $\Delta_{1}(z)$ takes the form
\begin{equation}
\frac{\Delta_{1}(z_{1}>0)}{\Delta_{0}}\approx8\sqrt{\frac{\pi}{7\zeta(3)}}\cos\delta\varphi\left(\frac{T_{c}}{\Delta_{0}}\right)^{3/2}\left[\sum_{n=0}^{\infty}\frac{1}{2n+1}\sin\theta_{0}^{\prime}(\omega_{n})\right]\exp\left(-z_{1}\frac{\Delta_{0}}{T_{c}}\frac{\sqrt{7\zeta(3)}}{\pi^{2}}\right).\label{eq:AppendixSGLDelta1Ans}
\end{equation}
Similarly the way we derived Eq.~\eqref{eq:AppendixSGLDelta1Ans1},
we can estimate the phase of the OP $\varphi_{1}(z)$
[see Eq.~\eqref{eq:AppendixSVarphi1AnsDep}]:
\begin{equation}
\varphi_{1}(z_{1}>0)-\varphi_{1}(+0)\approx2\sin\delta\varphi\frac{L_{11}(0)}{L_{20}(0)}\intop_{-\infty}^{+\infty}\frac{dk}{2\pi}\frac{1-\cos kz}{k^{2}}\approx
\frac{8}{\pi^{3/2}}z_{1}\sin\delta\varphi\left(\frac{T_{c}}{\Delta_{0}}\right)^{1/2}\sum_{n=0}^{\infty}\frac{1}{2n+1}\sin\theta_{0}^{\prime}(\omega_{n}).\label{eq:AppendixVarphi1GLAns}
\end{equation}

Now we find the $J_{2}^{(1,2)}(\Delta\varphi)$ corrections to the
current. In the considered limit $L/\xi^{\prime}\gg1$ one can use
Eqs.~\eqref{eq:AppendixLongJ21} and~\eqref{eq:AppendixLongJ22}.
In the calculation of $J_{2}^{(1,2)}(\Delta\varphi)$ we consider
the contributions of the S and S$^\prime$ regions separately and will take
into account only the main contributions with respect to $\Delta_{0}\ll T_{c}$
and $L/\xi^{\prime}\gg1$.

The S$^\prime$ region part in the formula for current $J_{2}^{(1)}(\Delta\varphi)$
is proportional to $\Delta_{0}/T$ and should be neglected, while
in the S part the first expression under the integral gives main contribution
of the order $\sim\!\!1$ (see Eq.~\eqref{eq:Ldefine} and take into account that the first term in $S(k)$ under the integral~\eqref{eq:AppendixLongJ21} converges on the scale $k\sim\left(L_{30}(0)/L_{20}(0)\right)^{1/2}\sim\left(\Delta_{0}/\omega\right)^{1/2}\ll1$].
Carrying out similar estimates for $J_{2}^{(2)}(\Delta\varphi)$ we
come to the conclusion that the main contribution comes from the first
term of the function $S(k)$ in the S region. Therefore,
\begin{multline}
J_{2}^{(1)}(\Delta\varphi)\approx-\frac{\sin\left(\Delta\varphi/2\right)\sgn\left[\cos\left(\Delta\varphi/2\right)\right]}{eR_{\text{int}}}\alpha\frac{\Delta_{0}^{2}}{\pi\Delta_{0}^{\prime}}\intop_{-\infty}^{+\infty}dk\frac{L_{1,1}^{2}(0)}{k^{2}L_{2,0}(0)+L_{3,0}(0)}\\
\approx-\frac{\alpha_{1}\sin\left(\Delta\varphi/2\right)\sgn\left[\cos\left(\Delta\varphi/2\right)\right]}{eR_{\text{int}}}\frac{16\pi}{\sqrt{7\zeta(3)}}\frac{T_{c}^{2}}{\Delta_{0}^{\prime}}\left[\sum_{n=0}^{\infty}\frac{1}{2n+1}\sin\theta_{0}^{\prime}(\omega_{n})\right]^{2},\label{eq:AppendixGLSLongJ21Ans}
\end{multline}
\begin{equation}
J_{2}^{(2)}(\Delta\varphi)\approx\frac{\sin\Delta\varphi}{2eR_{\text{int}}}\alpha\frac{\Delta_{0}}{\pi}\intop_{-\infty}^{+\infty}dk\frac{L_{1,1}^{2}(0)}{k^{2}L_{2,0}(0)+L_{3,0}(0)}\approx
\frac{\alpha_{1}\sin\Delta\varphi}{eR_{\text{int}}}\frac{8\pi}{\sqrt{7\zeta(3)}}\frac{T_{c}^{2}}{\Delta_{0}}\left[\sum_{n=0}^{\infty}\frac{1}{2n+1}\sin\theta_{0}^{\prime}(\omega_{n})\right]^{2}.\label{eq:AppendixGLSLongJ22Ans}
\end{equation}

\subsubsection{\texorpdfstring{$T\to T_{c}<T_{c}^{\prime},\ L\ll\xi^{\prime}$}{}}

In this limit, our estimates regarding the contribution of the S region
to the response for the current are preserved, but now we cannot neglect
the contribution of the S$^{\prime}$ region in $J_{2}^{(1)}(\Delta\varphi)$
given by Eqs.~\eqref{eq:AppendixShortJ21Ans}. As for $J_{2}^{(2)}(\Delta\varphi)$
the contribution of the S$^{\prime}$ region is of the order $\sim\Delta_{0}^{2}\xi^{\prime}/L$,
which we neglect in comparison with S region contribution $\sim T_{c}^{2}/\Delta_{0}$.
We obtain 

\begin{multline}
J_{2}^{(1)}(\Delta\varphi)\approx-\frac{\sin\left(\Delta\varphi/2\right)\sgn\left[\cos\left(\Delta\varphi/2\right)\right]}{eR_{\text{int}}}\left\{ \alpha\frac{\Delta_{0}^{2}}{\pi\Delta_{0}^{\prime}}\intop_{-\infty}^{+\infty}dk\frac{L_{1,1}^{2}(0)}{k^{2}L_{2,0}(0)+L_{3,0}(0)}\right.\\
+\left.\alpha^{\prime}\frac{\Delta_{0}^{\prime2}}{\Delta_{0}}\frac{2\xi^{\prime}}{L}\left[\frac{\left(L_{1,1}^{\prime}(0)-L_{3,1}^{\prime}(0)\right)^{2}}{L_{3,0}^{\prime}(0)}+L_{1,2}^{\prime}(0)-L_{3,2}^{\prime}(0)\right]\right\} \\
\approx-\frac{\sin\left(\Delta\varphi/2\right)\sgn\left[\cos\left(\Delta\varphi/2\right)\right]}{eR_{\text{int}}}\left\{ \frac{16\pi}{\sqrt{7\zeta(3)}}\alpha_{1}\frac{T_{c}^{2}}{\Delta_{0}^{\prime}}\left(\sum_{n=0}^{\infty}\frac{1}{2n+1}\sin\theta_{0}^{\prime}(\omega_{n})\right)^{2}\right.\\
\left.+4\alpha_{1}^{\prime}\Delta_{0}\frac{\xi_{1}^{\prime}}{L}\left[\frac{\left(\sum_{n=0}^{\infty}\frac{1}{2n+1}\sin\theta_{0}^{\prime}(\omega_{n})\cos^{2}\theta_{0}^{\prime}(\omega_{n})\right)^{2}}{\sum_{n=0}^{\infty}\sin^{3}\theta_{0}^{\prime}(\omega_{n})}+\sum_{n=0}^{\infty}\frac{1}{(2n+1)^{2}}\sin\theta_{0}^{\prime}(\omega_{n})\cos^{2}\theta_{0}^{\prime}(\omega_{n})\right]\right\},\label{eq:AppendixGLSShortJ21Ans}
\end{multline}

\begin{align}
J_{2}^{(2)}(\Delta\varphi) & \approx\frac{8\pi}{\sqrt{7\zeta(3)}}\frac{\alpha_{1}\sin\Delta\varphi}{eR_{\text{int}}}\frac{T_{c}^{2}}{\Delta_{0}}\left(\sum_{n=0}^{\infty}\frac{1}{2n+1}\sin\theta_{0}^{\prime}(\omega_{n})\right)^{2}.\label{eq:AppendixGLSShortJ22Ans}
\end{align}

\subsubsection{\texorpdfstring{$T\to T_{c}^{\prime}<T_{c},\ L\gg\xi^{\prime}$}{}}

In this case the answers for OP absolute value $\Delta_{1}^{\prime}(y)$
and phase $\varphi_{1}^{\prime}(y)$, and the current corrections
$J_{1,2}^{(1)}(\Delta\varphi)$ are given by formulas Eqs.~\eqref{eq:AppendixSGLDelta1Ans},~\eqref{eq:AppendixVarphi1GLAns}, and~\eqref{eq:AppendixGLSLongJ21Ans} with replacing all quantities from S region with quantities from S$^\prime$ region and vice
versa. In the case of $J_{2}^{(2)}(\Delta\varphi)$, it is also necessary
to take into account the contribution of the term $\alpha_{L}^{\prime}$:
\begin{gather}
\frac{\Delta_{1}^{\prime}(y_{2}<0)}{\Delta_{0}^{\prime}}\approx8\sqrt{\frac{\pi}{7\zeta(3)}}\left(\sum_{n=0}^{\infty}\frac{1}{2n+1}\sin\theta_{0}(\omega_{n})\right)\left(\frac{T_{c}^{\prime}}{\Delta_{0}^{\prime}}\right)^{3/2}\cos\delta\varphi\exp\left(y_{2}\frac{\Delta_{0}^{\prime}}{T_{c}^{\prime}}\frac{\sqrt{7\zeta(3)}}{\pi^{2}}\right),\label{eq:AppendixSPrGLLongDelta1Ans}\\
\varphi_{1}^{\prime}(y_{2}<0)-\varphi_{1}^{\prime}(-0)\approx\frac{8}{\pi^{3/2}}y_{2}\sin\delta\varphi\left(\frac{T_{c}^{\prime}}{\Delta_{0}^{\prime}}\right)^{1/2}\sum_{n=0}^{\infty}\frac{1}{2n+1}\sin\theta_{0}(\omega_{n}),\label{eq:AppendixSPrGLLongVarphi1Ans}\\
J_{2}^{(1)}(\Delta\varphi)\approx-\frac{16\pi}{\sqrt{7\zeta(3)}}\frac{\alpha_{2}^{\prime}\sin\left(\Delta\varphi/2\right)\sgn\left[\cos\left(\Delta\varphi/2\right)\right]}{eR_{\text{int}}}\frac{T_{c}^{\prime2}}{\Delta_{0}}\left[\sum_{n=0}^{\infty}\frac{1}{2n+1}\sin\theta_{0}(\omega_{n})\right]^{2},\label{eq:AppendixSPrGLLongJ21Ans}\\
J_{2}^{(2)}(\Delta\varphi)\approx\frac{\sin\Delta\varphi}{eR_{\text{int}}}\left[\sum_{n=0}^{\infty}\frac{1}{2n+1}\sin\theta_{0}(\omega_{n})\right]^{2}\left(\frac{8\pi}{\sqrt{7\zeta(3)}}\frac{\alpha_{2}^{\prime}T_{c}^{\prime2}}{\Delta_{0}^{\prime}}-\frac{4\alpha_{L}^{\prime}T_{c}^{\prime}}{\pi}\right).\label{eq:AppendixSPrGLLongJ22Ans}
\end{gather}

\subsubsection{\texorpdfstring{$T\to T_{c}^{\prime}<T_{c},\ L\ll\xi^{\prime}$}{}}

In this limit the junction is short $L\ll\xi^{\prime}$, thus we can
use Eqs.~\eqref{eq:AppendixShortDeltaAns},~\eqref{eq:AppendixShortVarphi1Ans},~\eqref{eq:AppendixShortJ21Ans}, and~\eqref{eq:AppendixShortJ22Ans}.
Implying $\Delta_{0}^{\prime}\ll T_{c}^{\prime}$ we obtain
\begin{gather}
\frac{\Delta_{1}^{\prime}(y_{2}<0)}{\Delta_{0}^{\prime}}\approx\frac{16\pi^{5/2}}{7\zeta(3)}\cos\delta\varphi\left(\frac{T_{c}^{\prime}}{\Delta_{0}^{\prime}}\right)^{5/2}\frac{\xi_{2}^{\prime}}{L}\sum_{n=0}^{\infty}\frac{1}{2n+1}\sin\theta_{0}(\omega_{n}),\label{eq:AppendixSPrGLShortDelta1Ans}\\
\varphi_{1}^{\prime}(y_{2})=\sin\delta\varphi\frac{\Delta_{0}}{\Delta_{0}^{\prime}}\left(y+\frac{L}{2\xi^{\prime}}\right),\label{eq:AppendixSPrGLShortVarphi1Ans-1}\\
J_{2}^{(1)}(\Delta\varphi)\approx-\frac{32\pi^{3}}{7\zeta(3)}\frac{\alpha_{2}^{\prime}\sin\left(\Delta\varphi/2\right)\sgn\left[\cos\left(\Delta\varphi/2\right)\right]}{eR_{\text{int}}}\frac{\xi_{2}^{\prime}}{L}\frac{T_{c}^{\prime3}}{\Delta_{0}\Delta_{0}^{\prime}}\left[\sum_{n=0}^{\infty}\frac{1}{2n+1}\sin\theta_{0}(\omega_{n})\right]^{2},\label{eq:AppendixSPrGLShortJ21Ans}\\
J_{2}^{(2)}(\Delta\varphi)\approx\frac{16\pi^{3}}{7\zeta(3)}\frac{\alpha_{2}^{\prime}\sin\Delta\varphi}{eR_{\text{int}}}\frac{\xi_{2}^{\prime}}{L}\frac{T_{c}^{\prime3}}{\Delta_{0}^{\prime2}}\left[\sum_{n=0}^{\infty}\frac{1}{2n+1}\sin\theta_{0}(\omega_{n})\right]^{2}.\label{eq:AppendixSPrGLShortJ22Ans}
\end{gather}

\subsubsection{\texorpdfstring{$T\to T_{c}=T_{c}^{\prime},\ L\gg\xi^{\prime}$}{}}

In this limit the answer for the current corrections $J_{2}^{(1,2)}(\Delta\varphi)$
is the sum of Eqs.~\eqref{eq:AppendixGLSLongJ21Ans}/\eqref{eq:AppendixGLSLongJ22Ans}
and~\eqref{eq:AppendixSPrGLLongJ21Ans}/\eqref{eq:AppendixSPrGLLongJ22Ans}, respectively. Taking into account $\Delta_{0}=\Delta_{0}^{\prime}$
we obtain
\begin{equation}
J_{2}^{(1)}(\Delta\varphi)\approx-\frac{\pi^{3}}{4\sqrt{7\zeta(3)}}\frac{\sin\left(\Delta\varphi/2\right)\sgn\left[\cos\left(\Delta\varphi/2\right)\right]}{eR_{\text{int}}}\left(\alpha_{1}+\alpha_{1}^{\prime}\right)\Delta_{0},\label{eq:GLLong1CorrCurrentAns}
\end{equation}
\begin{equation}
J_{2}^{(2)}(\Delta\varphi)\approx\frac{\Delta_{0}\sin\Delta\varphi}{eR_{\text{int}}}\left[\frac{\pi^{3}}{8\sqrt{7\zeta(3)}}\left(\alpha_{1}+\alpha_{1}^{\prime}\right)-\frac{\pi}{16}\alpha_{L}^{\prime}\frac{\Delta_{0}}{T_{c}}\right].\label{eq:GLLong2CorrCurrentAns}
\end{equation}
These results coincide with the previously derived expressions in Refs.~\cite{Kupriyanov1992,Osin2021} under the additional assumption $\alpha_{L}^{\prime}\Delta_{0}/T_{c}\ll\alpha_{1}$.

\subsubsection{\texorpdfstring{$T\to T_{c}=T_{c}^{\prime},\ L\ll\xi^{\prime}$}{}}

In this limit in order to find $J_{2}^{(1,2)}(\Delta\varphi)$ we
use Eqs.~\eqref{eq:AppendixJ21DeltaVarphiAns} and~\eqref{eq:AppendixJ2DeltaVarphiAns} 
due to the shortness of the junction. Taking into account $\Delta_{0}=\Delta_{0}^{\prime}\ll T_{c}$
we obtain
\begin{gather}
J_{2}^{(1)}(\delta\varphi)\approx-\frac{\pi^{5}}{14\zeta(3)}\frac{\alpha_{1}^{\prime}\sin\left(\Delta\varphi/2\right)\sgn\left[\cos\left(\Delta\varphi/2\right)\right]}{eR_{\text{int}}}\frac{\xi_{1}^{\prime}}{L}T_{c},\label{eq:GLShort1CorrCurrentAns}\\
J_{2}^{(2)}(\Delta\varphi)\approx\frac{\pi^{5}}{56\zeta(3)}\frac{\alpha_{1}^{\prime}\sin\Delta\varphi}{eR_{\text{int}}}\frac{\xi_{1}^{\prime}}{L}T_{c}.\label{eq:GLShort2CorrCurrentAns}
\end{gather}

\subsection{Applicability conditions of the perturbation theory}\label{AppendixApplicability}

When expanding $\Delta\ (\Delta^{\prime})$ into a series with respect
to the parameter $\alpha\ (\alpha^{\prime})$, we assumed that the correction
$\alpha\Delta_{1}$ is small compared to the bulk value of the order
parameter $\Delta_{0}$ (see Ref.~\cite{Osin2021}). As a result, we write a criterion for
the applicability of the approximations we have made
\begin{equation}
\alpha\frac{|\Delta_{1}(z=0)|}{\Delta_{0}}\ll1,\ \alpha^{\prime}\frac{|\Delta_{1}^{\prime}(y=0)|}{\Delta_{0}^{\prime}}\ll1.\label{eq:AppendixApplicabilityCriterion}
\end{equation}
Here we define the Ginzburg-Landau coherence length as follows:
\begin{gather}
\xi_{\text{GL}}^{2}(T)\equiv\frac{\pi D}{8T_{c}(1-T/T_{c})},\quad  \xi_{\text{GL}}^{\prime2}(T)\equiv\frac{\pi D^{\prime}}{8T_{c}^{\prime}(1-T/T_{c})},\label{eq:AppendixXiGlDef}\\
\alpha_{\text{GL}}\equiv\frac{g_{\text{int}}\xi_{\text{GL}}}{\sigma},\quad  \alpha_{\text{GL}}^{\prime}\equiv\frac{g_{\text{int}}\xi_{\text{GL}}^{\prime}}{\sigma}.\label{eq:AppendixAlphaGlDef}
\end{gather}

For the limit $T\to T_{c}\leq T_{c}^{\prime}$ the applicability criterion of
the perturbation theory in the S region extendable to arbitrary temperatures $T<T_{c}\leq T^{\prime}_{c}$ takes the form
\begin{gather}
\alpha_{1}\ll\left(\frac{\Delta_{0}}{T_{c}}\right)^{2}\left/\sum_{n=0}^{\infty}\right.\frac{1}{2n+1}\sin\theta_{0}^{\prime}(\omega_{n})\Leftrightarrow\alpha_{\text{GL}}\ll\frac{\Delta_{0}}{T_{c}}\left/\sum_{n=0}^{\infty}\right.\frac{1}{2n+1}\sin\theta_{0}^{\prime}(\omega_{n}),\ T_{c}<T_{c}^{\prime}\label{eq:AppendixSCritTcLessTc'}\\
\alpha_{1}\ll\frac{\Delta_{0}}{T_{c}}\Leftrightarrow\alpha_{\text{GL}}\ll1,\ T_{c}=T_{c}^{\prime},\label{eq:AppendixSCritTcEqTc'}
\end{gather}

For the limits $T\to T_{c}^{\prime}\leq T_{c}$, $L/\xi^{\prime}\ll1$ ($L/\xi^{\prime}\gg1$) the applicability of the perturbation theory in S$^\prime$ region extendable to arbitrary temperatures $T<T^{\prime}_{c}\leq T_{c}$ take the
form 
\begin{gather}
\alpha_{\text{GL}}^{\prime}\ll\left(\frac{\Delta_{0}^{\prime}}{T_{c}^{\prime}}\right)^{2}\frac{L}{\xi_{2}^{\prime}}\left/\sum_{n=0}^{\infty}\right.\frac{1}{2n+1}\sin\theta_{0}(\omega_{n}),\ T_{c}^{\prime}<T_{c},\ L\ll\xi^{\prime},\label{eq:AppendixSPrTc'LessTcShortCrit}\\
\alpha_{\text{GL}}^{\prime}\ll\frac{\Delta_{0}^{\prime}}{T_{c}^{\prime}}\frac{L}{\xi_{2}^{\prime}},\ T_{c}^{\prime}=T_{c},\ L\ll\xi^{\prime},\label{eq:AppendixSPrTc'EqTcShorCrit}\\
\alpha_{\text{GL}}^{\prime}\ll\frac{\Delta_{0}^{\prime}}{T_{c}^{\prime}}\left/\sum_{n=0}^{\infty}\right.\frac{1}{2n+1}\sin\theta_{0}(\omega_{n}),\ T_{c}<T_{c}^{\prime},\ L\gg\xi^{\prime},\label{eq:AppendixSPrTc'LessTcLongCrit}\\
\alpha_{\text{GL}}^{\prime}\ll1,\ T_{c}=T_{c}^{\prime},\ L\gg\xi^{\prime}\label{eq:AppendixSPrTc'EqTcLongCrit}
\end{gather}
\end{widetext}
The case of $L/\xi^{\prime}\gg1$ is similar to the results
above up to replacing $\Delta_{0},\ \alpha_{\text{GL}}$ with $\Delta_{0}^{\prime},\ \alpha_{\text{GL}}^{\prime}$
and vice versa.

\bibliography{biblio}

\end{document}